\documentclass[fleqn,usenatbib]{mnras}

\usepackage{newtxtext,newtxmath}

\usepackage[T1]{fontenc}

\DeclareRobustCommand{\VAN}[3]{#2}
\let\VANthebibliography\thebibliography
\def\thebibliography{\DeclareRobustCommand{\VAN}[3]{##3}\VANthebibliography}

\usepackage{graphicx}	
\usepackage{amsmath}	
\usepackage{hyperref}

\newcommand{\kpc}{\mathrm{kpc}}
\newcommand{\Myr}{\mathrm{Myr}}
\newcommand{\Msun}{\mathrm{M}_\odot}
\newcommand{\dex}{{\rm dex}}
\newcommand{\lbrac}[1]{\left(#1\right)}

\newcommand{\citeb}[1]{\citeauthor{#1} \citeyear{#1}}

\title[Nature of clumps using deep learning]{The nature of giant clumps in high-z discs: a deep-learning comparison of simulations and observations}

\author[O. Ginzburg et al.]{
Omri Ginzburg,$^{1}$\thanks{E-mail: omry.ginzburg@mail.huji.ac.il}
Marc Huertas-Company,$^{2,3,4,5,6}$
Avishai Dekel,$^{1,6}$
Nir Mandelker,$^{7,8,9,1}$\newauthor
Gregory Snyder,$^{10}$
Daniel Ceverino,$^{11,12}$
Joel Primack$^{13}$
\\
$^{1}$Racah Institute of Physics, The Hebrew University, Jerusalem 91904 Israel\\
$^{2}$LERMA, Observatoire de Paris, PSL Research University, CNRS, Sorbonne Universités, UPMC Univ. Paris 06,F-75014 Paris, France\\
$^{3}$Univeristé de Paris, 5 Rue Thomas Mann - 75013, Paris, France\\
$^{4}$Departamento de Astrofísica, Universidad de La Laguna, E-38206 La Laguna, Tenerife, Spain\\
$^{5}$Instituto de Astrofísica de Canarias, E-38200 La Laguna, Tenerife, Spain\\
$^{6}$SCIPP, University of California, Santa Cruz, CA 95064, USA\\
$^{7}$Kavli Institute for Theoretical Physics, University of California, Santa Barbara, CA 93106, USA\\
$^{8}$Department of Astronomy, Yale University, PO Box 208101, New Haven, CT, USA\\
$^{9}$Heidelberger Institut fŁur Theoretische Studien, Schloss-Wolfsbrunnenweg 35, D-69118 Heidelberg, Germany\\
$^{10}$Space Telescope Science Institute, Baltimore, MD, United States\\
$^{11}$Departamento de Fisica Teorica, Facultad de Ciencias, Universidad Autonoma de Madrid, Cantoblanco, 28049 Madrid, Spain\\
$^{12}$CIAFF, Facultad de Ciencias, Universidad Autonoma de Madrid, 28049 Madrid, Spain\\
$^{13}$Physics Department, University of California, Santa Cruz, Santa Cruz, CA 95064, USA
}

\date{Accepted XXX. Received YYY; in original form ZZZ}

\pubyear{2015}

\begin{document}
\label{firstpage}
\pagerange{\pageref{firstpage}--\pageref{lastpage}}
\maketitle

\begin{abstract}
We use deep learning to explore the nature of observed giant clumps in high-redshift disc galaxies, based on their identification and classification in cosmological simulations. Simulated clumps are detected using the 3D gas and stellar densities in the VELA zoom-in cosmological simulation suite, with $\sim \!\! 25\ \!\rm{pc}$ maximum resolution, targeting main sequence galaxies at $1\!<\!z\!<\!3$. The clumps are classified as long-lived clumps (LLCs) or short-lived clumps (SLCs) based on their longevity in the simulations. We then train neural networks to detect and classify the simulated clumps in mock, multi-color, dusty and noisy HST-like images. The clumps are detected using an encoder-decoder convolutional neural network (CNN), and are classified according to their longevity using a vanilla CNN. Tests using the simulations show our detector and classifier to be $\sim80\%$ complete and $\sim80\%$ pure for clumps more massive than $\sim10^{7.5}\Msun$. When applied to observed galaxies in the CANDELS/GOODS S+N fields, we find both types of clumps to appear in similar abundances in the simulations and the observations. LLCs are, on average, more massive than SLCs by $\sim 0.5\ \dex$, and they dominate the clump population above $M_{\rm c}\gtrsim 10^{7.6}\ \Msun$. LLCs tend to be found closer to the galactic centre, indicating clump migration to the centre or preferential formation at smaller radii. The LLCs are found to reside in high mass galaxies, indicating better clump survivability under supernova feedback there, due to clumps being more massive in these galaxies. We find the clump masses and radial positions in the simulations and the observations to agree within a factor of two.
\end{abstract}




\section{Introduction}

Star forming galaxies, residing in dark matter haloes of masses ${\sim 10^{12}\ \Msun}$ at redshifts $1\!<\!z\!<\!3$, are observed to be turbulent, rotation-supported extended discs with a typical rotation-to-dispersion ratios $V/\sigma\!\sim\!4$ \citep{Genzel2006,Forster2006,Forster2009,Simons2017,Forster2018}.

The images of high-redshift discs typically show giant clumps. The clumps are especially pronounced in rest-frame UV and ${\rm H_\alpha}$ emission maps, which serve as star-formation tracers \citep{EE06,Genzel2006,Genzel2008,Forster2009,Forster2011,Wuyts2012,Wuyts2013,G15,Shibuya2016,G18,HuertasCompany2020}. Clumps are also detected in interferometric submilimeter observations. Along with strong gravitational lensing, these observations are able to probe resolutions of $30-100\ \rm{pc}$, and observe dense emitting dust regions in $\rm{CO}$ maps at redshifts $z\!\sim\!1-2$ (\citealp{Swinbank2010, Dessauges2019}; cf. \citealp{Ivison2020}).

Observed giant clumps are found to host of order $\sim\!10\!-\!20\%$ of the galaxy's star formation rate (SFR) in the redshift range ${1\lesssim z\lesssim3}$ \citep{Wuyts2012,Wuyts2013,G15}. Clump masses are, on the other hand, still uncertain. Traditional SED fitting to clump fluxes, applied to clumps detected in low resolution observations, indicate clumps with stellar masses as high as $10^{10}\ \Msun$ \citep{G18,Zanella2019}, an order-of-magnitude more massive than simulated clumps. However, due to the limited resolution of these observations and projection effects, these clump masses tend to be over estimated, and need to be corrected to match high resolution simulations and observed clumps in lensed galaxies \citep{Cava18,HuertasCompany2020,Meng2020}.

\smallskip Detection of clumps in observations is usually performed either visually \citep[e.g.][]{Elmegreen2007,Forster2011} or automatically \citep[e.g.][]{G15,Zanella2019} in rest frame UV or optical, as well as CO or $\rm{H_\alpha}$ emission maps. The automatic detection is usually done by smoothing the image to extract bright residual objects. Recently, we used deep learning-based detectors that were trained to identify blobs in smooth S\'{e}rsic profile backgrounds, which proved to be more efficient than other automated clump detection schemes, with $\sim\!10\!-\!20\%$ increase in completeness (\citeb{HuertasCompany2020}). The deep learning method was also able to rapidly analyze complete samples of CANDELS galaxies, greatly increasing the number of galaxies studied.

\smallskip The sizes of the largest clumps in the simulations turn out to be in the ball park of a few hundred parsec, with the compact ones smaller than 100 pc. Their study thus requires simulations with resolution better than $\sim\! 50\ \rm{pc}$ in the disc \citep{Agretz09,Genel12,M14,Tamburello2015,M17,Oklopcic2017}.
\cite{M17} studied clumps in the VELA simulations \citep{Ceverino2014,Zolotov2015}, and found that the contribution to the galaxy mass is only $0.1\%\!-\!3\%$, as compared to the $1\%\!-\!30\%$ found in observations \citep{Genzel2006,G18,Meng2020,HuertasCompany2020}. The SFR contribution is also found to be lower in VELA, around $\sim3\!-\!6\%$ independent of redshift.

\smallskip The giant clumps could have formed ex-situ, namely be the remnants of small galaxies that merged with the disc. Alternatively, they could form in-situ, due to instabilities within the disc. The common understanding of the formation and evolution of in-situ clumps is that they arise from violent disc instability (VDI), in turbulent discs with a Toomre $Q$ parameter slightly smaller than unity \citep*{Toomre64,DekelSariCeverino09}. In this scenario, clump masses and sizes should scale with those of their host disc, particularly the cold component, mostly consisting of cold gas and young stars. The properties of clumps in simulations of isolated discs as well as fully cosmological simulations are fairly consistent with this picture \citep{Noguchi99,Immeli04,Immeli04b,BournaudElemegreen09,M14,Oklopcic2017,M17,Meng2020}. It is worth mentioning, though, that there are indications in cosmological simulations for proto-clump regions where the value of the Toomre $Q$ parameter is significantly above unity. This suggests an alternative element in the clump formation scenario, where compressive modes of turbulence initiate the clump formation \citep{Inoue2016}.

\smallskip Once collapsed and forming stars, the clump may either survive intact for many dynamical times or be disrupted in a few dynamical times. It is expected that the primary process responsible for clump disruption is stellar and supernova feedback, as deduced from comparisons of simulations with different feedback strength \citep[][Ceverino et al. in prep., Ginzburg et al. in prep]{M17}, while clump longevity is also affected by the dependence of clump properties on gas fraction \citep{Fensch2020}. Thus the longevity of clumps thus holds important information about feedback processes in galaxies \citep{Mayer2016}, which is one of the major unknowns in galaxy evolution. The determination whether clumps are long lived (LLCs) or short lived (SLCs) can help us develop better feedback models. The clump longevity is indeed an issue of debate in the simulations, perhaps reflecting the different subgrid recipes assumed for feedback in the different simulations. The high resolution, VELA simulations show a clear bi-modality in clump lifetimes, with a correlation between longevity and the mass of the clump (with a critical baryonic mass of $\sim\!10^8\ \Msun$) and the potential well depth \citep{M17}. On the other hand, other cosmological simulations find that most clumps are short lived, and are disrupted before virialization due to supernova feedback and winds \citep{Genel12,Tamburello2015,Oklopcic2017}. Besides the use of different hydro-gravitational codes, these simulations differ in the subgrid recipes for star formation and feedback, which is likely the main reason for the different results concerning clump longevity. Indeed, the outflow strength in these simulations have a typical mass loading factor $\eta\equiv\dot{M}_{\rm out}/SFR \sim 4-10$ and outflow velocities of the order of $400-700\ {\rm km\cdot s^{-1}}$ \citep{Genel12,Muratov2015}. In contrast, the mass-loading factor for long-lived massive clumps in VELA, as seen in Fig. 13 of \cite{M17}, is of order unity or less.  The quoted values for the clumps, which dominate the star formation, are expected to be in the same ballpark as the global mass-loading factor. A mass loading factor of order unity and less is consistent with observations of stellar and supernova outflows of ionized gas \citep{Forster2019}.

\smallskip In addition, the clump longevity has implications on the dynamical and structural evolution of the galaxy. Several studies, both theoretical \citep*{DekelSariCeverino09, DekelBurkert2014} and using simulations \citep{Noguchi99,Immeli04,Immeli04b,Bournaud07,Elmegreen2008,Ceverino2010,Bournaud2014} found that if clumps live long enough, they migrate inwards within a few orbital times and build up a galactic bulge, while gaining mass, growing their stellar component, and becoming older and redder. The clump migration is also responsible for stirring up turbulence in the disk, by turning potential energy into turbulent energy \citep{DekelSariCeverino09,Cacciato2012}. Color gradients in observations suggest similar scenario of clump migration.
The observed radial gradients in the clump population within the disc may indeed suggest clump migration \citep{Genzel2008,Forster2011,Shibuya2016,G18}. Alternatively, the gradients may reflect preferred formation location.
We note, however, that once a galaxy develops an extended gaseous ring about a massive central mass, the inward migration of the clumps in the ring is significantly suppressed  (\citeb{DekelRings}). 

\smallskip The observational studies of clump longevity so far were limited to stellar-population age estimates via SED fitting and ad-hoc models of star-formation history \citep{Forster2011,G18}. These age estimates carry very large uncertainties. In this work, we aim to study the longevity of clumps in an innovative way, using deep learning and full 3D knowledge of clumps using simulations.

 \begin{figure*}
    \centering
    \includegraphics[width=2\columnwidth]{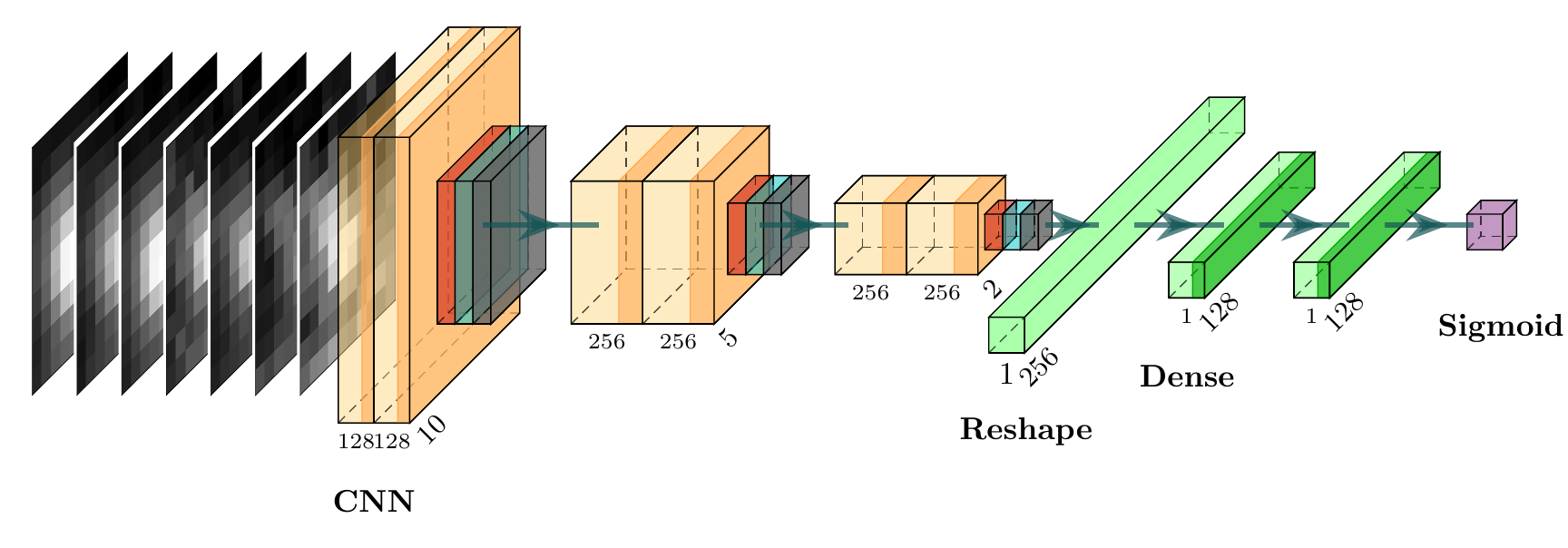}
    \caption{Illustration of the CNN architecture. A multi-channel small stamp is provided to the machine, and an output $P\in[0,1]$ is provided, representing the probability to be long lived. Light orange shaded bands represent a 2D convolutional layer, while the opaque dark orange band represent a ReLU activation. Red blocks represent a $2\times2$ max pooling layer, cyan blocks represents a dropout layer and a gray box represents a batch normalization layer. Light green transparent blocks represent dense layer, and an opaque green band represents a ReLU activation. The horizontal numbers represent the depth of the layer, and the slant numbers represent the spatial dimension in the CNN part and the length of the dense layers in the dense part. The illustration was done using the {\tt PlotNeuralNet} tool (\href{https://github.com/HarisIqbal88/PlotNeuralNet}{https://github.com/HarisIqbal88/PlotNeuralNet}).}
    \label{fig:Cnn}
\end{figure*}

 \smallskip With the ever increasing data available for astrophysical studies, machine learning is becoming a powerful tool for studying and analyzing the data.
 In recent years, deep neural networks (NNs) achieved world-wide fame and recognition due to their ability to out-perform many previously used machine learning algorithms, as well as their opening a new realm of deep generative models, which provides the ability to learn and sample complex distributions (e.g. \citeb{Margalef2020} for an astrophysical context). Thanks to technological advancements in parallel computing (GPUs, TPUs), such models are rapidly improving and becoming more readily available. However, the excellent performance comes with a price. Deep NNs are highly complicated models, consisting of a large number of free parameters and complex relations between them. Because of that, deep NNs are considered `black boxes' in the sense of interpretability - one usually cannot determine how the machine produced the output it produced, although some methods exist to interpret them (e.g. \citeb{IntGrad}, see \citeb{HuertasCompany2018} for an astrophysical context). In this study we do not attempt to interpret the brain of the machine in a systematic way.
 \begin{figure*}
    \centering
    \includegraphics[width=2\columnwidth]{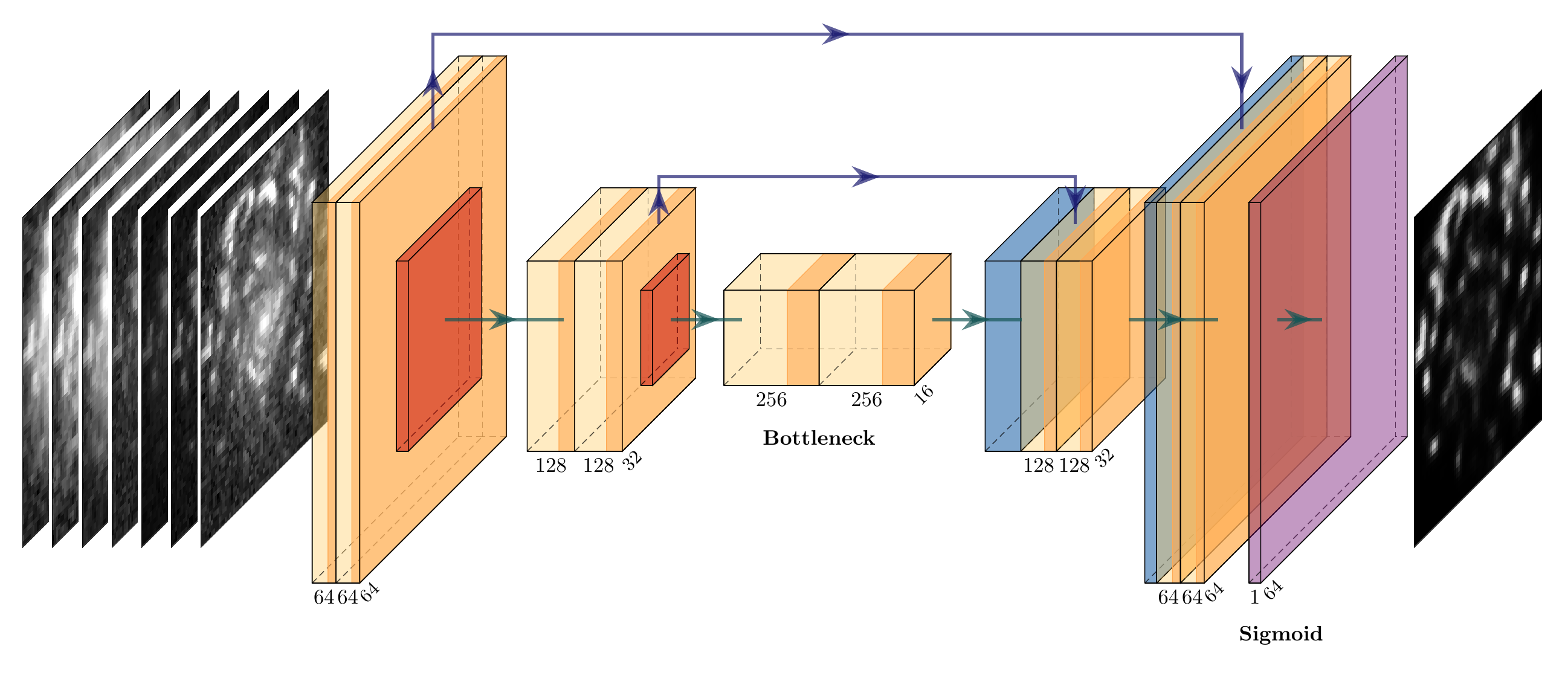}
    \caption{Illustration of the UNet architecture. A multi-channel image is provided to the machine, and a grayscale image is produced. Light orange transparent bands represent a 2D convolutional layer, while the opaque dark orange band represent a ReLU activation. Red blocks represent a $2\times2$ max pooling layer, and a blue block represents a 2D transposed convolution. In each block, the horizontal number represents the number of convolutional filters used, and the slant number represents the spatial dimension of the block. The last layer is a pixel-wise sigmoid activation, producing the grayscale map. The green arrows are the direct links, and the blue arrows represent the `skip links` of the UNet. }
    \label{fig:Unet}
\end{figure*}

Deep learning has been successfully used to compare simulated and observed galaxies, and test different theoretical models. \cite{HuertasCompany2018} trained a deep NN to classify CANDELS galaxies into their phase of evolution with respect to the blue nugget (BN) phase \citep*[pre, during and post BN;][]{Zolotov2015,DekelLapinerDubois2019} as defined in the VELA simulations. They identified the BNs with similar abundances as in simulations, and revealed a favorable mass for BNs in the observed sample. \cite{Margalef2020} used a deep generative model to find morphological outliers in CANDELS using Horizon-AGN simulation suite as a training set. They found that the Horizon-AGN simulations do not reproduce the smallest galaxies and high S\'{e}rsic index galaxies. \cite{Ferreira2020} trained a deep NN to classify galaxies to their merger phase using IllustrisTNG. They used the trained models to compute the cosmological merger rates of CANDELS galaxies, with good agreement with pair counts estimates.

In this paper, we apply deep-learning methods to the study of the nature of giant clumps, following the logic of learning the clump properties from the simulations in order to find and classify them in the observed sample. The machine will be trained on mock HST images of the VELA galaxies to detect and classify clumps into longevity classes (see \S\ref{sec:in_situ}, \S\ref{sec:training_set} below) and will be applied to real CANDELS galaxies.

The paper is organized as follows. In \S\ref{sec:networks}, we describe the network architectures used. The reader is referred to Appendix A of \cite{Hausen2019} for the deep-learning terminology used. A reader who is not interested in the details of the DL methods can skip this section without hurting the ability to understand the results. In \S\ref{sec:clump_cat} we present the VELA simulations and the derived clump catalog, as well as the procedure of generating the mock images. In \S \ref{sec:training_set} we describe the training strategy of our networks. In \S\ref{sec:pipeline_performance} we evaluate the performance of our pipeline. In \S\ref{sec:observations} we run our pipeline on CANDELS GOODS-S+N in order to study the nature of the observed giant clumps. In \S\ref{sec:discussion} we discuss on the possible features our machine learns, as well as the effects of supernova feedback on clump longevity. Finally, in \S\ref{sec:conc} we summarize our results.

\section{Network architectures}\label{sec:networks}
The main tool we use in this study is a pipeline that extracts and classifies clumps in 2D HST-like images. The pipeline consists of two parts. The first part consists of providing a 2D image of a galaxy, and retrieving all of its clumps. The second part consists of classifying small stamps of clumps to their longevity class. To achieve this goal, we use two types of deep NNs, described in the following subsections.

\subsection{Vanilla CNN - Classification}\label{sec:CNN}
For the classification of the clumps, we use a simple, vanilla, deep Convolutional Neural Network (CNN). Traditional NNs consist of neurons and weighted links (`synapses') between them. Each neuron has a value which depends on the input and the weights connecting to it. On the other hand, in CNNs, the input is usually a 2D greyscale or multi-channel image. The network is built by `convolutional layers', which consist of fixed-size filters. Each filter represents a convolution operator performed on the output of the preceding layer. These convolution operators allow the network to learn spatial correlation between pixels and are translationally invariant. 
CNNs are often used in classification tasks as subnetworks inside a larger network. Traditionally, they are concatenated to a `fully connected' network, which is a classical NN consisting of neurons and links (although some use so-called `fully convolutional' networks for various tasks). 

\smallskip The network architecture we use is illustrated in Figure~\ref{fig:Cnn}.
Each block in our CNN classifier consists of two convolutional layers, each followed by a ReLU activation function (which introduces non linearities, in order to learn complex models), and a $2\times2$ max pooling layer at the end. After each max pooling a dropout layer is added with a dropout rate of $0.7$. This unusually large rate is used in order to prevent overfitting due to our small dataset. Each block ends with a batch normalization layer. After three blocks, the output from the last layer is passed to a fully connected network with two dense layers, each followed by a ReLU activation. The output layer consists of a single neuron with a sigmoid activation, translating the network output to a a value between zero and one, representing a probability.
\subsection{UNet - Detection}
For detecting clumps, we use the UNet architecture (\citeb{Unet}). The UNet architecture, a variant of the encoder-decoder architecture, was designed to perform semantic segmentation tasks, originally for biomedical imaging. The traditional encoder-decoder architecture is comprised of two paths. The first, a contracting path, designed for extracting important features from the image using traditional convolutional and pooling layers. The second, an expanding path, designed to use the lower dimensional data representation of the input to reconstruct the image to its original dimensions. The particularity of the UNet is the so called `skip links' - the contracting and expanding paths have the same number of levels, and activation maps in the expanding path are concatenated with the output of the corresponding level in the contracting path. The method of skip links provides the machine additional information on how to reconstruct the image. The UNet has already been used in other astrophysical studies \citep{Hausen2019,HuertasCompany2020,Hong2020,Boucaud2020}.

The network architecture is illustrated in Figure~\ref{fig:Unet}. We use a basic implementation of the UNet. Each block in the contracting path consists of two 2D convolutional operators, each followed by a ReLU activation, and a $2\times2$ max pooling layer at the end. Each block in the expanding path consists of a 2D transposed convolution operator (which is a learnable upsampling operator, allowing the increase of the dimensions of the data), followed by two convolutional operators, each followed by a ReLU activation. The final layer is a $1\times1$ convolution layer followed by a sigmoid activation.  After each block in the contracting (expanding) path, the output (input) is downsampled (upsampled) by a factor of two, while the number of filters increases (decreases) by a factor of two. The output from each contracting block is concatenated after the transposed convolution in the corresponding expanding block.

\section{Simulation dataset}\label{sec:dataset}
The training set of our models is derived solely from the VELA simulations. In \S\ref{sec:vela} we give a short summary of the main properties of the VELA simulations, and refer the reader to appendix A of \cite{M17} and \cite{DekelDisk} for more details. In \S \ref{sec:clump_finder} we review the clump finding method developed in \cite{M14} and \cite{M17}. In \S\ref{sec:in_situ}, we describe the different types of clumps found in VELA. In \S\ref{sec:candelized} we describe how the mock VELA images were generated. Lastly, in \S \ref{sec:training_set}, we describe how we put everything together to build up our training set.
\begin{figure*}
    \centering
    \includegraphics[width=2\columnwidth]{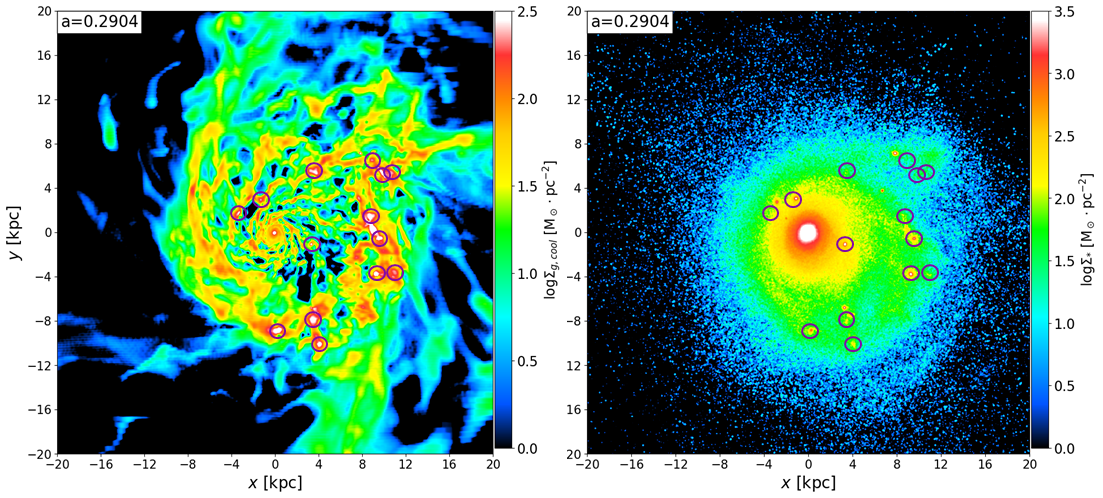}
    \caption{Face-on projections of the cool gas ($T<1.5\times 10^4\ K$; left) and stellar (right) densities of one of the VELA galaxies ($V07$), at redshift $z\!=\!2.44$. The dark circles represent several detected clumps using the 3D detector (see text). We can see several clumps that are pronounced in the gaseous map, while they have little or no stellar component.}
    \label{fig:clumps_in_sim}
\end{figure*}
\subsection{3D Clump catalog}\label{sec:clump_cat}
\subsubsection{Summary of VELA simulations}\label{sec:vela}
The VELA3 (henceforth 'VELA') suite consists of 34 high resolution, zoom-in, cosmological simulations using the ART code (\citeb{Kravtsov1997},\citeb{Ceverino2009},\citeb{Ceverino2014}). Most of the galaxies reach $z\sim 1$, with maximum spatial resolution between physical $17-35\ \mathrm{pc}$ at all times. The dark matter halo masses at $z=2$ are in the range of ${10^{11}\ \Msun - 10^{12}\ \Msun}$. The galaxies were selected at $z=1$ from a low resolution, dark matter only run, to have halo virial mass $\sim10^{11-12} \Msun$ and to not be experiencing an ongoing major merger (mass ratio $>\!1\!:\!4$) at that time. This eliminated less than $10\%$ of the haloes in the simulated mass and redshift ranges, which were likely to end up as ellipticals at low redshifts. The simulation outputs are evenly spaced in $a$, the cosmological scale factor, with $\Delta a = 0.01$. This corresponds to $\Delta t\in (80,160)\ \Myr$, typically $\sim 120\ \Myr$.

\smallskip The code consists of various cooling mechanisms, such as atomic hydrogen and helium cooling, metal and molecular line cooling. Heating by a spatially uniform and time dependent photoionizing UV background, with partial self-shielding, is also included. Besides the cooling and heating mechanisms, the code also incorporates star formation, stellar mass loss, metal enrichment of the ISM and stellar feedback. Supernovae and stellar winds are incorporated by local injection of thermal energy, and radiation pressure feedback is also implemented \citep{Ceverino2009,Ceverino2010,Ceverino2012,Ceverino2014}.

The stellar and supernova feedback implemented in the VELA simulations tend to be on the low-strength side compared to existing cosmological simulations. This in turn produces lower values for the gas fraction of galaxies and higher values of the stellar-to-halo mass ratio. 

The strength of the feedback is clearly important for the evolution of clumps, especially their disruption or survival, which is one of our major issues here. For example, \cite{Moody14} and \cite{M17} studied the effect of radiation pressure (RP) feedback on the formation and evolution of clumps by analyzing a counterpart simulation of the same VELA initial conditions, run without RP. While the overall effect on the entire clump population was minor, \cite{M17} found differences in LLC abundances between simulations with and without RP. Furthermore, preliminary examination of VELA6, a new re-simulations of the VELA galaxies with stronger kinetic feedback model, motivated by the effects of clustered supernova following \cite{Gentry2017}, which better match the predicted stellar-to-halo mass relation than the simulations studied here (Ceverino et al., in prep.), reveals a large decreases in the abundance of LLCs (Ginzburg et al., in prep.). It is not clear whether this increase of feedback strength overestimates the effects on clumps. \cite{Oklopcic2017} studied clumps in a single galaxy from the FIRE simulation suite in the redshift interval $1<\!z\!<2.2$, with comparable resolution to the VELA galaxies used here but with a different strong feedback prescription. They found that, although the galaxy tends to be clumpy for a gigayear timescale, the median age of individual clumps was only $\sim20\ \Myr$. In other words, all their clumps were short-lived. In an Appendix, they confirmed that two other FIRE simulations at $z\!\sim\!2$ had similar clump properties. 

The VELA simulations have sufficient resolution to resolve clumps. However, the price of a high resolution is a small number of galaxies available for analysis. This poses a challenge to deep learning practices, due to the very small training set with respect to common deep learning standards. Nevertheless, we learned from \cite{HuertasCompany2018} that with proper fine-tuning a similar deep learning method can be applied successfully to such a small training set in order to extract very useful results. This encouraging success motivates our attempt here. 

\subsubsection{3D clump detection}\label{sec:clump_finder}
The high spatial resolution present in the VELA simulations ($17\!-\!35\ \rm{pc}$), is crucial for studying clumps, as the typical clump sizes are $\sim 180\ \rm{pc}$ in radius \citep{M17,Oklopcic2017}. Below, we outline our method for detecting and characterizing clumps in the simulations.

Clumps are detected using the full 3D distributions of gas and stars in each galaxy. We begin by interpolating these onto a uniform grid with a cell-size of $70\ {\rm pc}$ using the cloud-in-cell technique. The grid is then smoothed with a (spherical) Gaussian filter with a full-width-at-half-maximum of $\min\!{\lbrac{0.5R_d,2.5\ \kpc}}$, where $R_d$ is the radius incorporating $85\%$ of the cold disc mass, including cold gas ($T<1.5\cdot10^4\ K$) and young stars (younger than $100\ \Myr$), within $15\%$ of the halo's virial radius. The density residual above the smoothed density field, $\rho_0$, is then calculated, $\delta = \rho/\rho_0 - 1$, once for the cold component (cold gas and young stars) and once for the stellar component, and the overdensity field at each point is set to be the maximum between the two. Clumps are defined as connected regions of at least eight grid cells, with density residual $\delta>10$. After defining the clumps, gas, stars and dark matter are deposited into the clump cells. An example of a clumpy galaxy in our simulations is presented in Figure~\ref{fig:clumps_in_sim}.

The clumps are, in general, not spherical. However, we define a clump radius as the radius of a sphere centred at the clump centre of mass, with the same volume as the clump: $\left({4\pi}/{3}\right)r_c^3 = (70\ \mathrm{pc})^3N$, where $N$ is the number of cells belonging to the clump. For a minimum of eight cells, this translates to a minimal clump diameter of $\sim\! 180\ \mathrm{pc}$.

Clumps are tracked between snapshots using their stellar particles (as VELA does not have tracer particles for gas). Only clumps which contain at least 10 stellar particles were tracked. For each clump in a given snapshot, the preceding snapshot is searched for progenitors - clumps which have contributed at least $25\%$ of their stellar particles to the target clump. If a clump has more than one progenitor, the most massive one is considered the main progenitor, while the others are considered as having merged. In case the preceding snapshot has no progenitors, the process continues to the previous-to-preceding snapshot and so on, until a progenitor is found or the initial snapshot of the simulation is reached.

\smallskip Each clump is assigned a `clump time', $t_{\rm c}$, which is an approximation for the time since the clump has formed. This measure will allow us to define the longevity of the clump. In practice, the clump time is defined as follows. At the first snapshot the clump is identified, at cosmic time $t_i$, the mass-weighted-mean stellar age of the clump is calculated, $\left<age\right>$. If $\left<age\right>\leq t_i-t_{i-1}\ (\sim 120\ \Myr)$, we set $t_{\rm c} = \left<age\right>$, otherwise we set $t_{\rm c}=0$. At each succeeding snapshot where the clump is identified, the cosmic time between snapshots is added to $t_{\rm c}$. The clump lifetime, $t_{\rm c,max}$, is defined as $t_{\rm c}$ at the last snapshot in which the clump was identified.
\begin{figure*}
    \centering
    \includegraphics[scale=0.6]{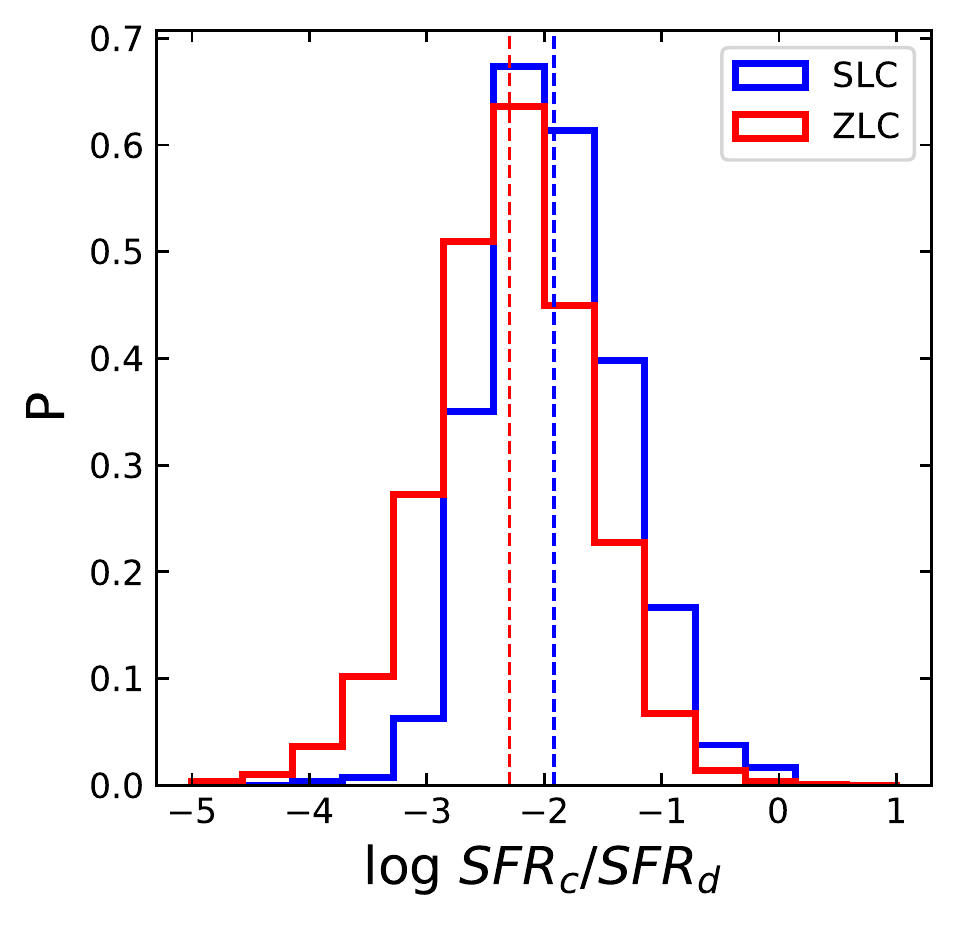}
    \includegraphics[scale=0.6]{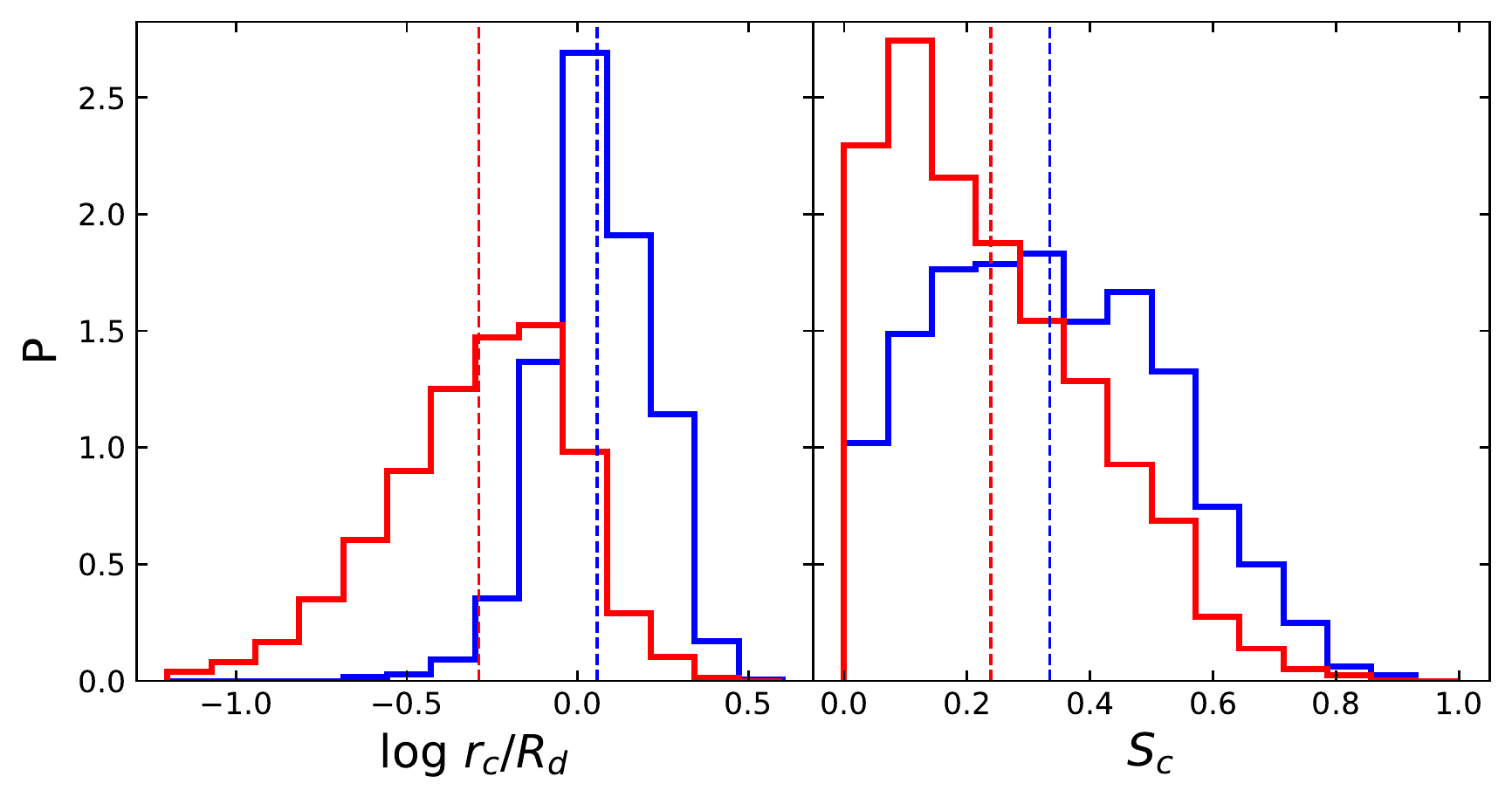}
    \caption{Distribution of various properties of SLCs vs ZLCs. Each histogram is normalized to a distribution. The ZLCs in this figure are selected to have $SFR_c>0$. In the left we show the distribution of the normalized SFR of the clumps. Though bit shifted, the ZLC distribution is very similar to the SLC distribution. In the middle panel we show the distribution of the positions of the clump normalized to the disc radius. We can see that these star-forming ZLCs are closer the the centre than SLCs. In the right panel we show the distribution of the shape parameters of the clumps (see text). ZLCs tend to have a more filamentry shape than SLCs, possibly due to tides.}
    \label{fig:zlc_dist}
\end{figure*}

\subsubsection{In-situ clumps}\label{sec:in_situ}
\cite{M17} defined 3 main classes of clumps: bulge, in-situ and ex-situ clumps. A bulge clump is a clump whose centre resides within two grid cells from the galactic centre. Ex situ clumps are external merging galaxies, and are defined based on their dark matter overdensities (see discussion in \S 3.3.3 in \citeb{M17}). The rest of the clumps are in-situ clumps, formed within the disc by VDI due to Toomre instability, or perhaps other non-linear VDI, e.g. compressive modes of turbulence \citep{DekelSariCeverino09,Inoue2016}.

The ex-situ population comprises only $5\%$ of the off-centre clump sample, but they contribute $\sim\!40\%$ in mass. Unfortunately, for our study, the $1\!\!:\!\!20$ imbalance in the number of clumps, and the small absolute number of clumps we have make it difficult to classify whether clumps are in-situ or ex-situ using machine learning. 

\smallskip The in-situ clumps were further classified to three sub-classes, based on their longevity: short lived clumps (SLCs), long lived clumps (LLCs) and zero-life clumps (ZLCs). For each clump, we calculate the mass-weighted mean free-fall time over the lifetime of the clump, $t_{\mathrm{ff}} = \left<\sqrt{{3\pi}/{32G\rho}}\right>$. Then,
\begin{itemize}
    \item If $t_{\rm c,max} = 0$, the clump is defined as a $ZLC$.
    \item If $0<t_{\rm c,max} < 20t_{\mathrm{ff}}$, the clump is defined as a $SLC$.
    \item If $t_{\rm c,max}\geq 20t_{\mathrm{ff}}$, the clump is defined as a $LLC$.
\end{itemize}
The distribution of clump lifetimes in the simulations was found to be bi-modal, with a separation at $\sim20 t_{\rm ff}$, thus motivating the above classification. However, we note that $20 t_{\rm ff}$ roughly corresponds to the simulation timestep. 
By these definitions, potentially, LLCs can live for one snapshot if they are not very dense, and SLCs can live more than one snapshot if they are very dense. However, these are extreme minorities in each population, and therefore negligible statistically, namely, the SLCs typically disrupt between two consecutive snapshots, and the LLCs survive for more than one snapshot.

\smallskip ZLCs comprise $\sim\!75\%$ of the clump sample. \cite{M17} found them to be low mass, low density filamentry-shaped patches, with a virial parameter $\alpha_{\rm v} = a\sigma^2r_{\rm c}/GM_{\rm c}$ (where $a$ is a constant of order unity, $\sigma$ is the clump velocity dispersion, $r_{\rm c}$ and $M_{\rm c}$ are the clump radius and mass) much greater than the SLCs, and therefore excluded them from the analysis. Nevertheless, a closer examination of these clumps reveals that a subsample of them have a distribution of SFRs very similar to that of the SLCs. Figure~\ref{fig:zlc_dist} shows distributions of three clump properties for the ZLCs (red) and SLCs (blue). Only ZLCs with non-zero star formation (henceforth termed `sfZLCs') are included in these statistics. We show the distributions of clump SFR normalized to the underlying disc SFR (left), the clump galacto-centric distance normalized by the disc radius (centre), and a shape parameter (right, defined below). We can see that the sfZLCs have a similar distribution of disc-normalized SFRs, though the typical values are $~0.3\ \mathrm{dex}$ lower. On the other hand, the sfZLCs tend to be much closer to the galactic centre, with a broad tail going all the way to $\sim0.15R_d$. 

\smallskip As evident from our method for assigning clump age (see \S \ref{sec:clump_finder}) the defining property of ZLCs is that they are identified for only one snapshot, and have a mean stellar age greater than the time since the previous snapshot. Being closer to the centre, and since no attempt was done to remove unbound material (see \citeb{M17}), many ZLCs may, in fact, be real bound clumps that are contaminated by old background stars.

\smallskip The fact that ZLCs tend to be closer to the centre may also explain the differences in the shape parameter. The shape parameter is defined as $S_c = I_3/I_1$ where $I_1\!>\!I_3$ are the largest and smallest moments of inertia of the clump. $S_c\!\sim\!1$ corresponds to a round clump, while $S_c\!\ll\!1$ corresponds to an elongated structure. Being closer to the centre makes the ZLCs more prone to strong gravitational interactions with the hosting galaxy. Tidal effects become stronger, and clump deformation and disruption is more easily achieved. 
ZLCs can also be remnants of stripped LLCs, which happens close to the centre during their migration. If a clump forms new stars while losing most of its old stars due to stripping, then fewer than 25\% of the clump stars may still be present in the clump in the following snapshot. In such a case, the clump in the earlier snapshot would not be identified as a progenitor of the clump in the later snapshot. Due to the unknown nature of the ZLCs in our simulations, it is difficult to associate them to a certain longevity class (SLC or LLC) without thoroughly studying their formation, which is beyond the scope of this paper. The ZLCs are important for the evaluation of the purity of our detection scheme, as further discussed in \S\ref{sec:unet_perf}. We eventually do not include the ZLC class in the classification scheme, as further discussed in \S\ref{sec:cnn_training}.

\subsubsection{CANDELized images}\label{sec:candelized}
The VELA \textit{CANDELized} images\footnote{Available publicly at \href{https://archive.stsci.edu/prepds/vela/}{https://archive.stsci.edu/prepds/vela/}} were introduced in \cite{Snyder2015}, and were used in a variety of studies \citep{HuertasCompany2018,Simons2019,Mantha2019,HuertasCompany2020}. \cite{HuertasCompany2018} also used the mock images with deep learning, successfully classifying CANDELS galaxies into the phase of evolution with respect to the blue nugget phase, namely to pre-, during and post-BN phase \citep*{Zolotov2015,DekelLapinerDubois2019}. Here we briefly summarize the generation of the mock images, and refer the reader to the aforementioned papers for a more in depth explanation.

A spectral energy distribution is assigned to each stellar particle in the simulation based on its mass, age and metallicity. The dust density is assumed to be proportional to the metal density as read from the simulation, with a dust-to-metal ratio of $0.4$ and assuming a grain size distribution from \cite{Draine2007}. Radiative transfer is then performed by {\tt SUNRISE} \citep{Jonsson2006,Jonsson2010,JonssonPrimack2010} using Monte Carlo ray tracing - the energy emitted by every star particle at each wavelength is propagated through the ISM while being scattered or absorbed by dust grains until it exists the grid or enters one of the viewing apertures (cameras). The raw mock images are created by integrating the SED in each pixel over the spectral response function of the CANDELS ACS and WFC3 filters ($F606W,F435W,F850LP,F775W, F105W,F125W,F160W$). Images are then convolved with the corresponding HST point spread function. The raw images are produced without noise, which is added post-processing (see \S \ref{sec:unet_training}).

Nineteen different orientations were used for the mock images. Two cameras are fixed to the instantaneous angular momentum (AM) direction (`face-on') of the cold disc, two cameras are fixed to the `edge-on' direction, one camera is fixed to $45^{\mathrm{o}}$ of the AM direction. Three cameras are fixed to the $x,y,z$ directions of the simulation box. Four cameras are randomly selected at the initial snapshot, and applied to all subsequent snapshots. The last seven cameras are randomized in each snapshot. For the training and performance evaluation of the networks, we use images with projected minor-to-major axis ratio $b/a>0.5$ as produced by running {\tt GALFIT} \citep{Galfit} on the images. If {\tt GALFIT} does not converge, we use images whose camera direction is within $45^{\mathrm{o}}$ of the galaxy AM direction.

\subsection{Training set \& strategy}\label{sec:training_set}
The VELA simulation suite is not large enough to constitute a statistical sample. This is critical when using deep learning, as a relatively large fraction of the available dataset must be devoted for training, and cannot be used for analysis. To cope with this, we train multiple models. During the training of each model, one VELA galaxy, with all its snapshots and clumps, is left out of the training set, and is used as a test set. This trained machine is then evaluated on this test set.

As most LLCs in our suite reside in four galaxies, we train four \textbf{\textit{main}} models, which correspond to the four galaxies, as mentioned above. These four galaxies are clumpy, containing many SLCs and LLCs. Using these four models, we will analyze the observed CANDELS galaxies. We further train 13 \textbf{\textit{secondary}} models, which will not be used to analyze the observations. These, along with the four main models, will be used to construct a machine-made clump catalog of 17 VELA galaxies, which will be used to compare VELA to CANDELS in a consistent way.

\subsubsection{Clump detection}\label{sec:unet_training}
The images for the UNet were generated by cropping the CANDELized images to $64\times 64$ pixels. Images in the ACS filters were downsampled by a factor of two to match the resolution of the WFC3 filters, while conserving the flux. The dataset therefore consists of $\sim 8500$ images (after discarding all images that do not meet the alignment criteria, see above), each with dimensions $64\times64\times7$, where the last dimension is the `depth' of the images, i.e. the number of filters (channels) used.

\begin{figure*}
    \centering
    \includegraphics[width=2\columnwidth]{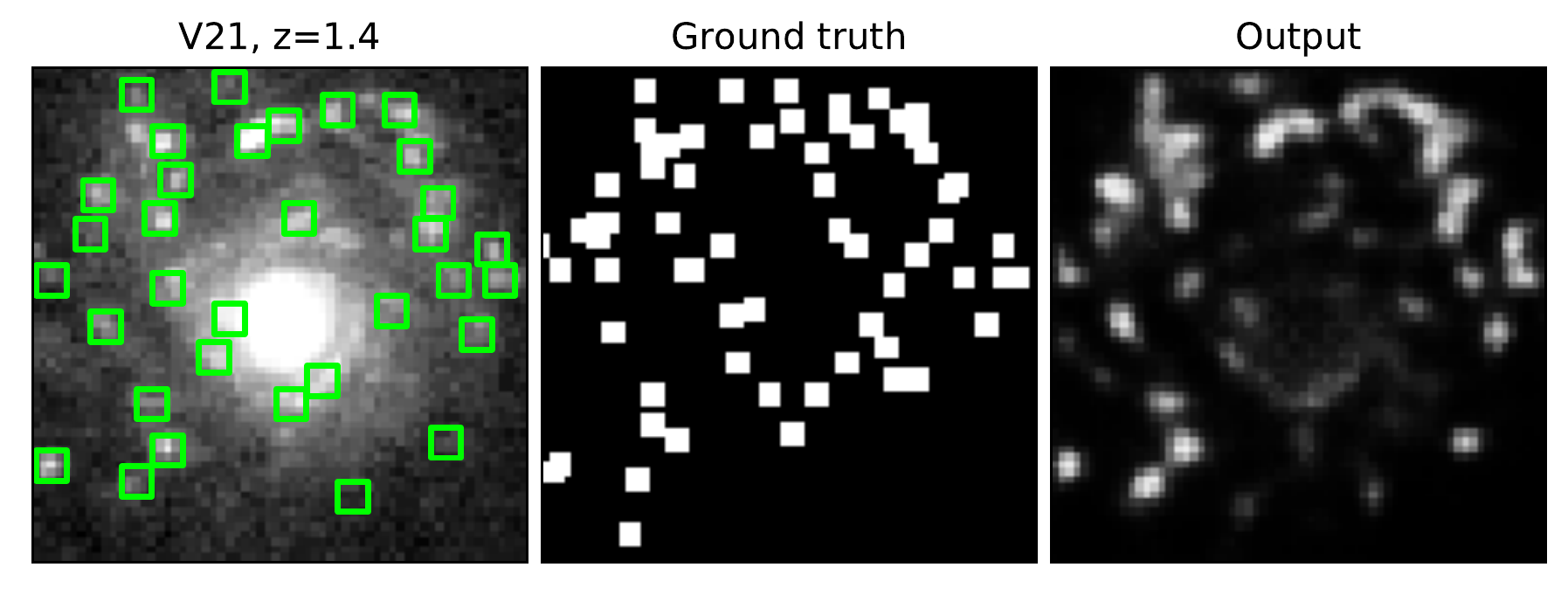}
    \caption{Example of the UNet+{\tt SExtractor} output. Shown is the result for one of the VELA galaxies at $z=1.4$, referring to one camera in the $F606W$ filter (out of seven channels used). \textit{\textbf{Left}}: The image of the galaxy in the $F606W$ filter, in one of the random orientations. The green squares indicate the positions of the clumps detected by the UNet+{\tt SExtractor} pipeline. \textit{\textbf{Middle}}: The `ground truth' map. Each $2\!\times\!2$ white square marks the projected positions of the 3D clumps onto the plane of the image. \textit{\textbf{Right}}: The output from the UNet, which is passed through {\tt SExtractor}. The value of each pixel is between $0$ and $1$, which represents the probability that a pixel belongs to a clump. We can see that the UNet+{\tt SExtractor} is able to detect many of the clumps found in the galaxy, while clumps that are clustered in a small region cannot be distinguished by the machine.}
    \label{fig:unet_example}
\end{figure*}

For the training of the UNet, we generate a collection of `ground truth' images. The UNet will be trained to transform a mock HST-like image into an image to resemble as much as possible the ground truth map (see Figure~\ref{fig:Unet}). The ground truth is a single channel, binary map, where each pixel's value is either one or zero. Pixels that belong to clumps are set to one. Ideally, we would determine the pixels which belong to a clump as those that fall within the 2D-projected 3D clump radius. However, as mentioned above, given the limited resolution of HST, most clump sizes fall below the observed resolution of HST of $0.5\ \kpc$. Therefore, at each projected position of a 3D clump, a square of $2\times 2$ pixels around the clump's position is set to pixel values of one. This corresponds to a physical side length of $\sim\!1\ \kpc$ in the redshift range considered here. An example is shown in Figure~\ref{fig:unet_example}.

As discussed in \S \ref{sec:in_situ}, many ZLCs have non-zero SFR, comparable to the SFRs of SLCs. This makes sfZLCs appear quite substantially in the bluer bands of the CANDELized images. As discussed below in \S \ref{sec:unet_perf}, even though trained to not detect  sfZLCs, the machine still finds these bright clumps appearing in the bluer bands. We therefore test our machine's performance with various thresholds on SFRs above which we include sfZLCs in our detection training set. The consequences of this are discussed in \S \ref{sec:unet_perf}.

\smallskip We train our network for a maximum of $4\times10^4$ iterations. Each training iteration consists of a `batch' of four images. Due to the small sample size, we use real-time data augmentation techniques. These techniques are used to effectively increase the size of the dataset by performing transformations on the images. In our case, the data augmentation is applied to both the images (channel-wide) and the ground truth maps. The data augmentations used here are as follows. During each iteration, we rotate the images in the batch by a multiple of $90^\mathrm{o}$\footnote{Any rotation which is not a multiple of $90^\mathrm{o}$ requires interpolating the subject image. Since our training labels are not rotationally-invariant, they must go through the same transformation. The fact that they are binary images causes the interpolation to produce bad results. We therefore did not use arbitrary angle rotations.} with $50\%$ probability, with $25\%$ probability for each multiple\footnote{Including zero, so a non-zero rotation is applied with probability of $37.5\%$}. Also, each image is reflected with a probability of $50\%$ about an axis of reflection (either horizontal or vertical) which is also chosen with $50\%$ probability. This effectively increases the data set by a factor of $16$, but with significant degeneracy among the images. However, these type of transformations force the machine to learn features of the detected objects beyond localization. To lift some of the degeneracy, we apply a non-conventional `patch-hiding' technique (\citeb{PatchHiding}). With every iteration, a random patch, of random size and location, from the $64\times 64$ image is set to zero. The idea behind this technique is to force the machine to learn more about the underlying features of the objects we want to detect, and less about their absolute position and relation to the overall image. This method increases the training set by an additional factor of $\!\sim\!10^3$, while introducing less degeneracy than other methods. 


\smallskip We make an effort to add realistic noise to our images. This has been demonstrated to significantly improve the performance of a CNN when attempting to identify mergers \citep{Bottrell2019}. To accomplish this, a random real CANDELS sky patch is added to each image (after all of the data augmentation procedures were applied) to imitate the correct noise level and correlation of real CANDELS observations.

\subsubsection{Clump classification}\label{sec:cnn_training}
Our building of the dataset for the classification CNN is quite unconventional. Recall that the pipeline procedure is (i) an image is passed through the UNet to detect clumps, and (ii) a squared stamp is cropped around each detected clump and passed through the CNN for the classification. In order to make this two-staged procedure consistent, our training set for the CNN is built upon the results of the UNet. Each image of each snapshot and camera is passed through one of the trained UNets to extract the clumps. Then an $n\times n\times 7$ stamp is cropped around each detected clump (see \S\ref{sec:unet_perf} for how $n$ was chosen). These small stamps are the images used for the classification scheme. 

We now define the classes used for the classification. For the remainder of this section, class $0$ will refer to SLCs and class $1$ will refer to LLCs (while ZLCs are ignored, see below). Because of the limited resolution of the observed images compared to the simulated clumps, we define the class of each stamp as follows. For each stamp, we consider all of the 3D clumps that fall inside the $n\times n$ square (in case there is more than one), and compute the mass-weighted average class, i.e.
$$
    \left<C\right>_s = \frac{\sum\limits_{i\in s}m_iC_i}{\sum\limits_{i\in s}m_i}
$$
where $i$ runs over all the 3D clumps in the square $s$, $m_i$ is the mass of the clump and $C_i$ is the class of the clump ($0$ or $1$) and $\left<C\right>_s$ is the mass-weighted average class of a square $s$. Then, the class of the image, $C_s$, is defined as
$$
C_s = \begin{cases}
0 & \left<C\right>_s < C_0 \\
1 & \left<C\right>_s \geq C_0\end{cases}
$$
for a choice of $C_0$. This is not very sensitive to the choice of $C_0$ because $\sim80\%$ of the images with $\left<C\right>_s>0$ have $\left<C\right>_s > 0.9$. We therefore choose $C_0=0.9$ to try and reduce contamination of LLCs by SLCs as much as possible.
\begin{figure*}
    \centering
    \includegraphics[width=\columnwidth]{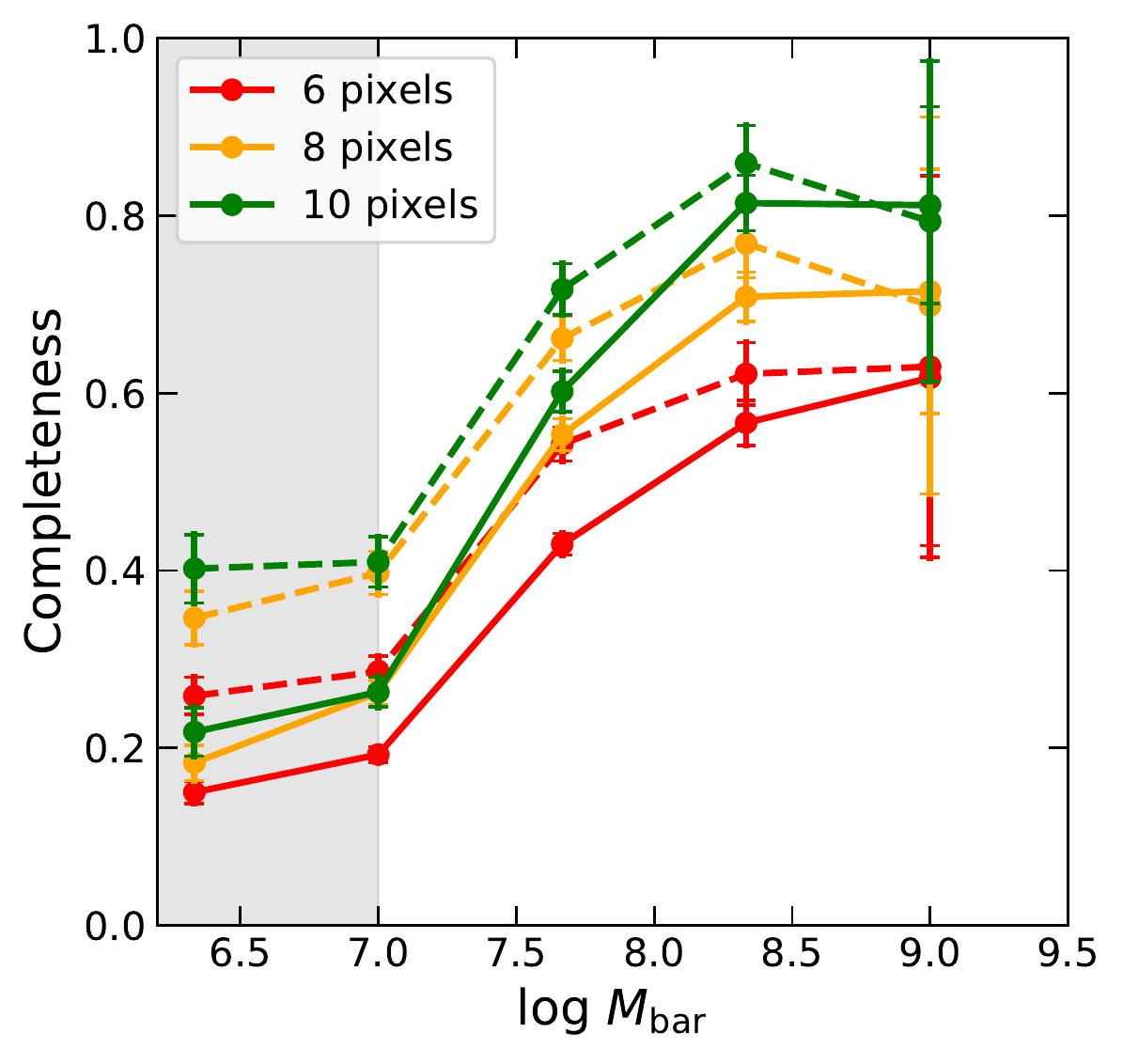}
    \includegraphics[width=\columnwidth]{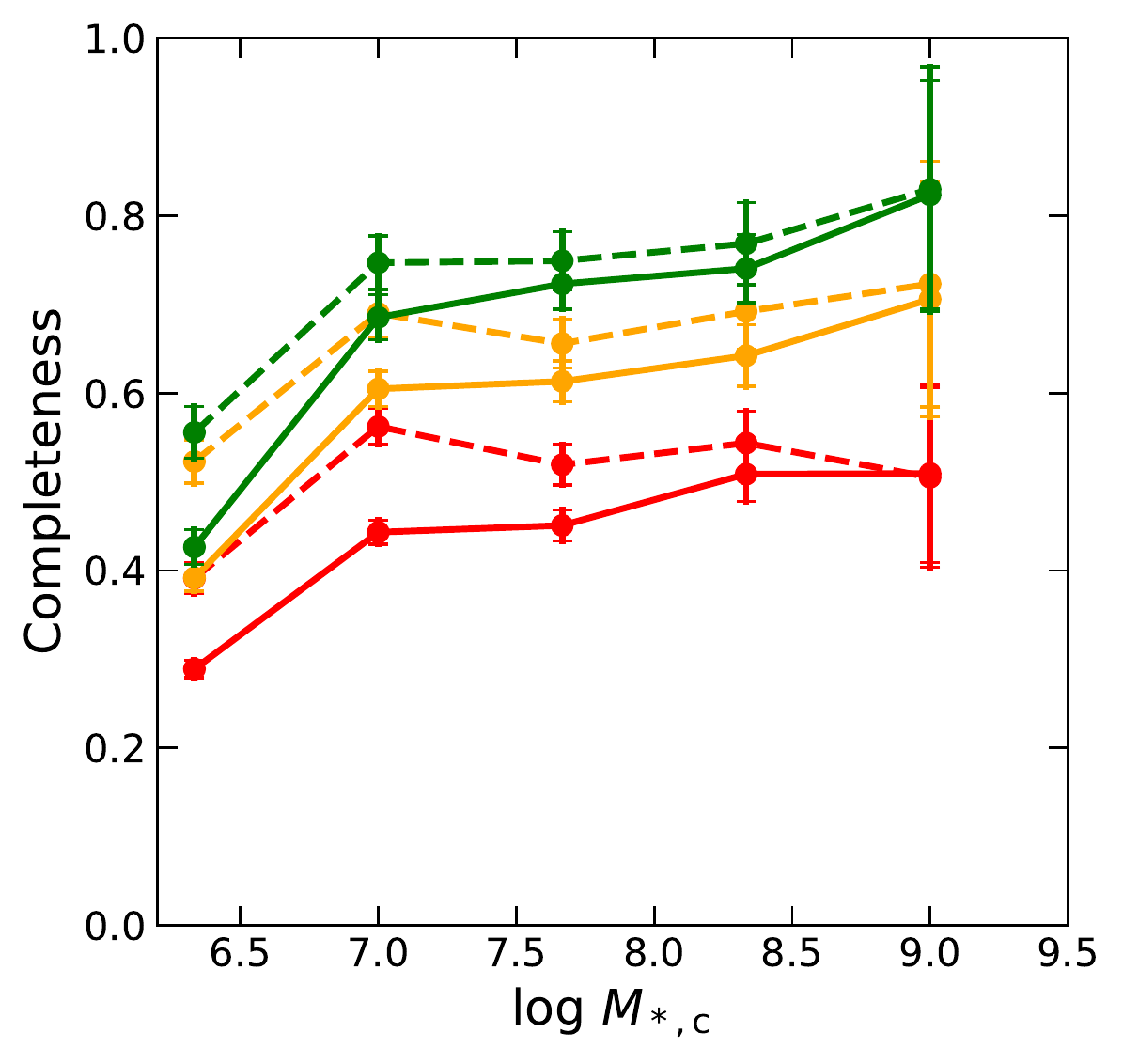}
    \caption{The completeness of our detection scheme. The completeness is shown as a function of the total baryonic mass (left) and stellar mass (right) in clumps within a square. Each colour refers to a different square size (see text for details). The solid lines refer to the networks trained including all of the sfZLCs, while the dashed lines refer to the networks trained including only the sfZLCs with $\rm{SFR_c}/\rm{SFR_d}>10^{-3}$. The gray area refers to the excluded completeness limit of the 3D clump detector of \protect\cite{M17}. Shown are the stacked results for the four main models we use. We see an increase of completeness with mass, which is more pronounced as a function of the baryonic mass. Including ZLCs has a stronger effect on completeness at lower masses.}
    \label{fig:completeness_unet}
\end{figure*}
\begin{figure}
    \centering
    \includegraphics[width=\columnwidth]{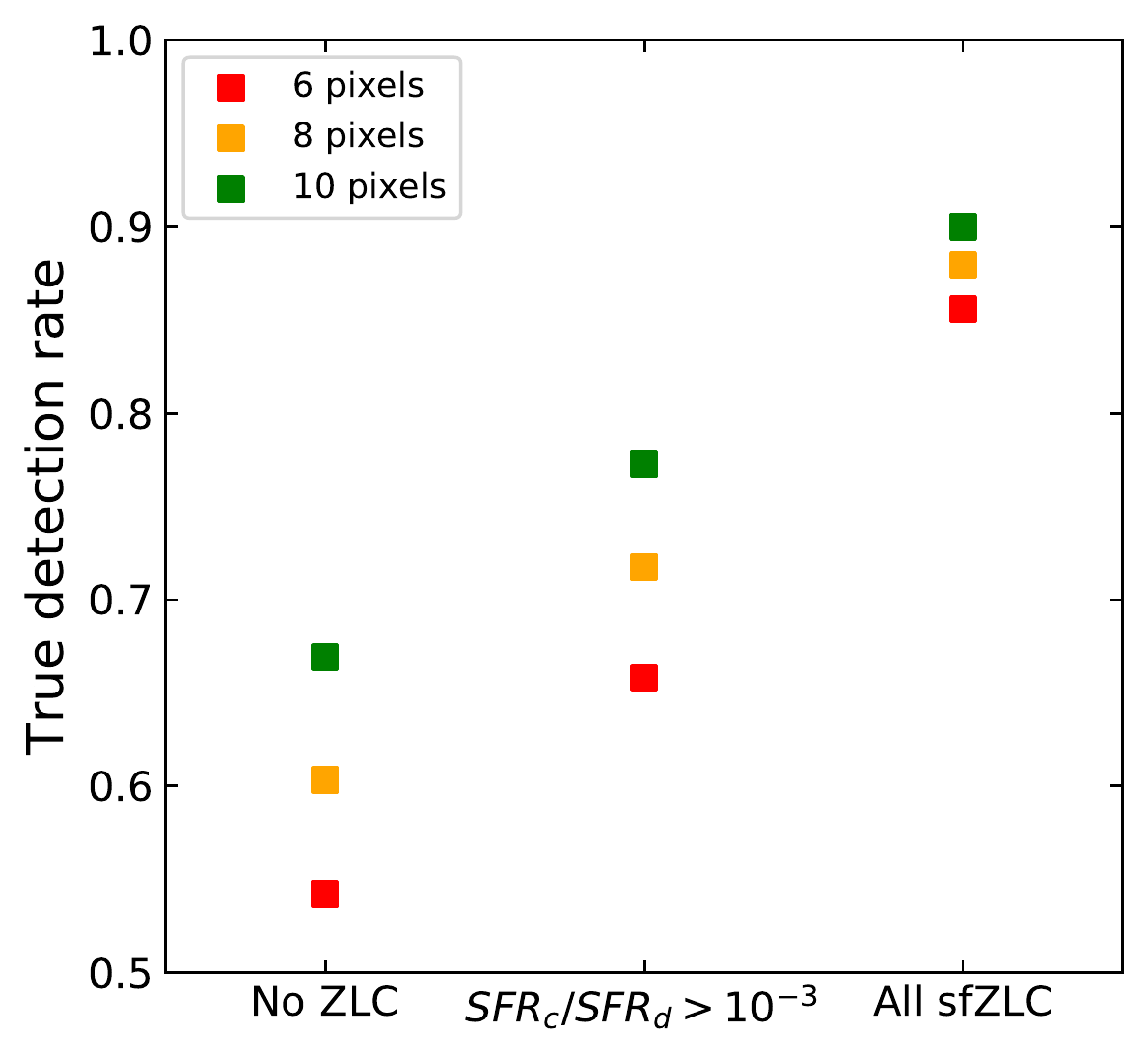}  
    \caption{The true detection rate of our detector as a function of the inclusion of ZLCs in the training set. Different colours are as in Figure~\protect\ref{fig:completeness_unet}. We can see how the true detection rate increases with the threshold put on the SFR of the sfZLCs, and with the square size, reaching $\sim\!90\%$ rate for square size of $10$ pixels, when including all of the sfZLCs.}
    \label{fig:false_detection}
\end{figure}
\begin{figure}
    \centering
    \includegraphics[width=\columnwidth]{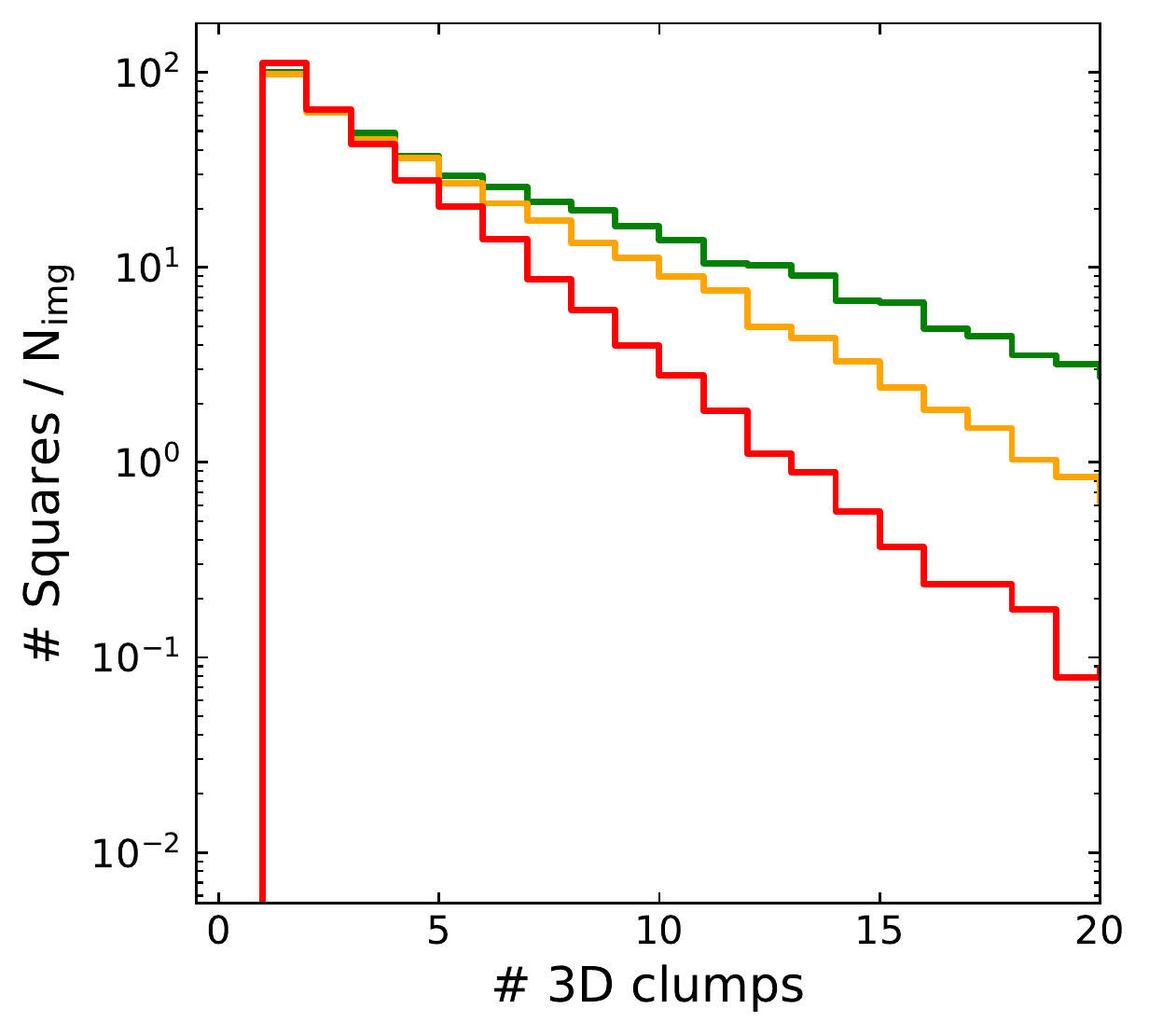}
    \caption{The distribution of the number of squares that host a given number of 3D clumps. The colours are the same as in Figure~\protect\ref{fig:completeness_unet}. Here we include all ZLCs, to get an upper bound. The vertical axis is normalized by the number of squares per image, in our case given by $\left(64 / \text{square size}\right)^2$. The median values of the number of 3D clumps are $2,3,4$ for square sizes $6,8,10$ pixels, respectively. The number of squares hosting less than five 3D clumps is roughly the same for all square sizes, and the distributions decrease exponentially. This fact suggests that using $10\!\times\!10$ squares around the detected clumps does not introduce a large loss in locality.}
    \label{fig:clumps_in_square}
\end{figure}

For $n=8\!-\!10$, the size of the dataset is usually $16\mathrm{k}-20\mathrm{k}$ stamps. The size of the dataset seems acceptable, but we note two drawbacks: (i) The images are very small, which increases the need for a larger dataset; (ii) the number of images corresponding to the two classes are very imbalanced, with a ratio of $1\!\!:\!\!5$ to $1\!\!:\!\!10$ in favor of the SLCs. To overcome the imbalance, we adopt the following training strategy. If $B$ is the batch size in each iteration, we draw from it a fraction $f\!\in\![0,1]$ such that $fB$ is an integer. Then, the batch is assembled by $fB$ random SLC images and $(1-f)B$ random LLC images. Using this strategy, the machine on average sees the same number of examples in each class, although it is more likely to see the same LLC multiple times. This is a form of an \textit{over-sampling} technique - the minority class is being over-sampled and is introduced to the machine multiple times. In our case, we choose $B=64$, and $f$ to be a random multiple of $1/8$.
We train for a maximum of $300\mathrm{k}$ batches. Besides that, we apply real-time data augmentation, including random rotations and reflections, as described above.

We wish to prevent the machine from simply learning the clump mass as the main distinguishing feature, as the mass may strongly correlate with clump longevity \citep{M17}.  For this, we assign to the clump in the H-band image a random magnitude in the range $19-24\ {\rm mag}$, and correct the rest of the channels accordingly.

Following the discussion in \S\ref{sec:in_situ}, we do not include any ZLCs in the calculation of the stamp's class, and we train the machine only on stamps that have at least one non-ZLC clump, while treating the ZLCs as background. Indeed, when we run stamps that contain only ZLCs through our trained classifier, we find that the machine is indecisive. Some of the ZLCs are classified LLCs with high probability, and some are classified as SLCs. It is therefore not possible to associate them to either class systematically, and we choose to ignore them.
    

\section{Pipeline performance}\label{sec:pipeline_performance}
\subsection{Clump detection}\label{sec:unet_perf}The output of the UNet is a probability map in which the pixel values, ranging from zero to one, represent the probability of a given pixel to be associated with a clump. In order to pin-down the positions of the clumps, we run the source extracting program {\tt SExtractor} \citep[which detects sources that are brighter than a threshold above their surrounding background;][]{Sextractor} on the output image from the machine, as we did in \cite{HuertasCompany2020}.
Figure~\ref{fig:unet_example} shows an example of the result of running the UNet+{\tt SExtractor} on one of the VELA galaxies at $z=1.4$. The galaxy is viewed with one of the random cameras, in the $F606W$ filter. Although the galaxy's structure is complex, the UNet+{\tt SExtractor} procedure is able to pin down many of the clumpy regions, while also showing a small number of failures. For example, the pipeline detected two clumps in the bottom right corner of the image, which do not show a counterpart in the ground truth map. Below, we quantify the machine's success to detect clumps in simulated galaxies.

The most natural way to evaluate the completeness of the detection scheme would have been to associate a detection with the closest 3D counter-part in the 2D image. This, however, turns out to be a very limited procedure. Due to the limited resolution of HST compared to VELA ($\sim 0.5\ \kpc$ and $17-35\ \mathrm{pc}$, respectively), a single detection in 2D can correspond to multiple 3D clumps. This can be seen in the ground truth map in Figure~\ref{fig:unet_example}, where we can see that several detected clumps (green squares) correspond to regions with many real clumps (white squares in the middle panel). Furthermore, as we showed in \cite{HuertasCompany2020}, resolution and projection effects can significantly influence the estimated clump properties (e.g. due to blending), possibly reaching an order-of-magnitude overestimation of clump masses. Similar conclusions were drawn in \cite{Meng2020}. Hence, we cannot assign a single 3D clump to a 2D detection in a consistent way.

\smallskip The way we evaluate the completeness of detection is as follows. We divide a $64\times64$ image of a galaxy into squares of a fixed size (e.g. $6,8$ or $10$ pixels). Each square is assigned a quantity which we use as a measure for completeness. For example, we can assign each square the total mass in 3D clumps that fall in that square, and we declare the detection successful when this mass is non-zero. Figure \ref{fig:completeness_unet} shows the completeness of detection as a function of clump baryonic mass and as a function of clump stellar mass, for three different square-side lengths. The completeness is computed for the results of all four main networks, stacked. We show the results for the networks trained with all sfZLCs (solid) and alternatively limited to intermediate sfZLCs ($SFR_c/SFR_d>10^{-3}$; dashed) . We can see, when using all of the sfZLCs, a clear systematic increase of completeness with baryonic clump mass, i.e. the completeness increases with baryonic clump mass, reaching $\sim75\%$ for squares of size $8$ pixels and $\sim80\%$ for squares of size $10$ pixels. A similar trend is visible for the completeness as a function of stellar mass, though the completeness seems to be flatter above $M_{\rm c}>10^7\ \Msun$.

\smallskip The full versus partial inclusion of ZLCs does not make a big difference to the completeness of detection, ranging from one to five per cent at high baryonic masses where the completeness is already high. The completeness with respect to stellar mass is even less affected. However, the threshold on the SFR of the included ZLCs does affect the amount of false detections with respect to our definition. To quantify these, we draw squares of the same sizes used above around each detection, and see if it falls close to a 3D clump. Figure \ref{fig:false_detection} shows the true detection rate of our detector, for a training set without any ZLCs, with intermediate sfZLCs and with all sfZLCs. We can see that the true detection rate increases dramatically with increasing SFR threshold, approaching $\sim\!90\%$ when all ZLCs are included.

\begin{figure*}
    \centering
    \includegraphics[width=1.5\columnwidth]{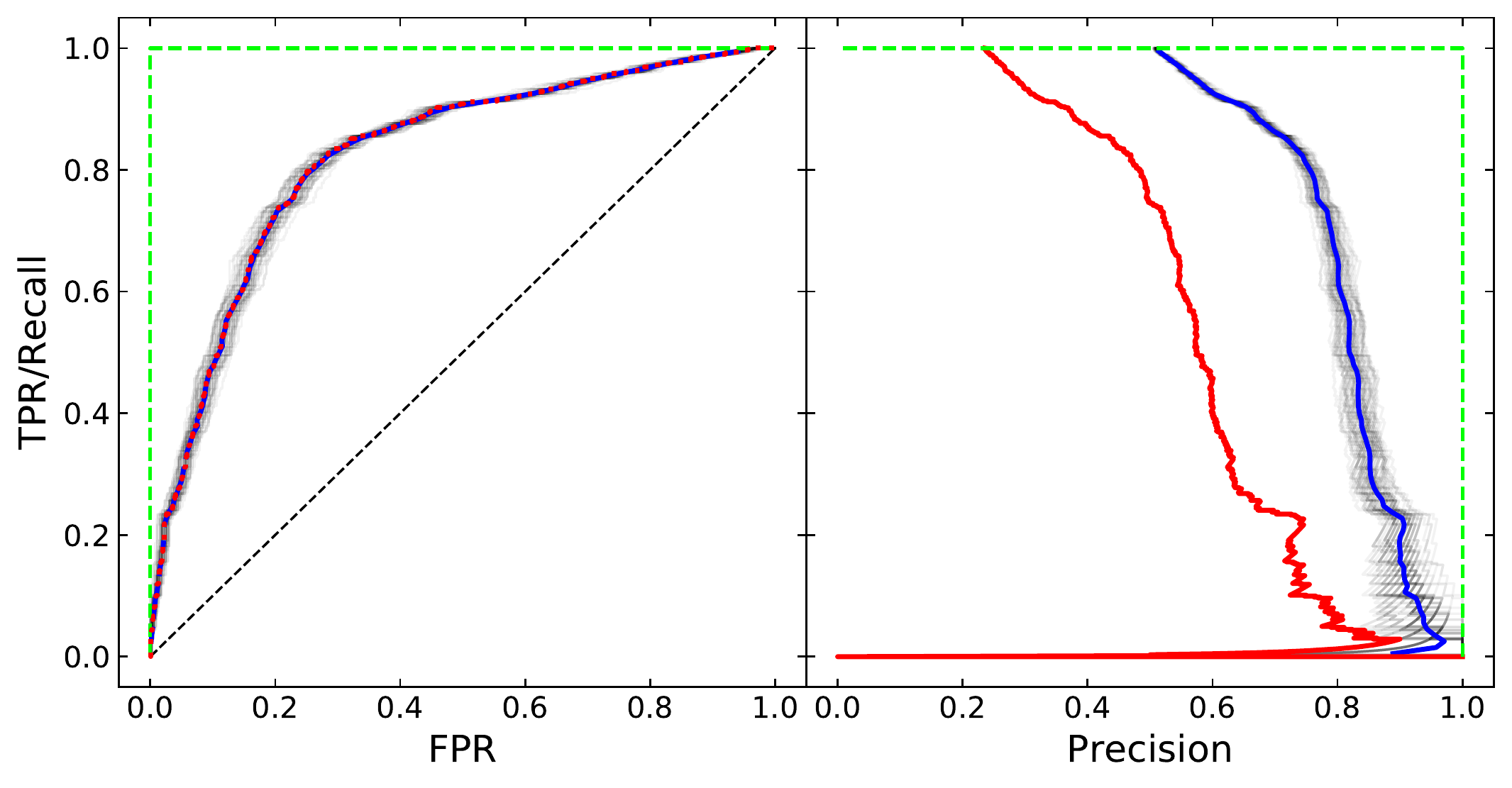}
    \caption{ROC (left) and precision-recall (right) curves for one of our classifiers. The green dashed lines in both panels are theoretical curves for an ideal classifier. The dashed black line in the left panel is the theoretical result for a completely random classifier. The red dots (left) and red curve (right) show the result on all of the test set. The black thin transparent curves in both panels show the results on randomly generated balanced subsets of the test set, and the blue curves show the mean curve (see text). The ROC curve is independent of the balance of the data set being validated, while the precision-recall curve is very sensitive, which is due to the large imbalance between the sets (see \S \protect\ref{sec:classification})}
    \label{fig:roc_prec}
\end{figure*}
\smallskip Our definition for completeness is affected by the square size we pick. Clearly, the larger the square the higher the probability for a detected clump and a 3D clump to fall inside the same square. On the other hand, a smaller square size has the advantage of making the detection more local. Besides being affected by the size of the region, the number of clumps within it is also affected by the limited resolution and projection effects. Figure \ref{fig:clumps_in_square} shows the distribution of the number of 3D clumps that fall inside squares of a given side. In order to take into account the automatic decrease of the number of squares when the square size is increased, we normalize the distribution by the number of squares per image. The difference in the number of squares with fewer than five clumps between square sizes of six and ten is less than a factor of two, and all distributions decay roughly exponentially. Given the natural uncertainty caused by the limited resolution, and in order to increase the completeness and true detection rate, we use squares of $10\!\times\!10$ pixels in our classification, corresponding to a radius of five pixels. This is almost the same size of aperture used when computing fluxes of clumps in other observational analyses \citep{Shibuya2016,G18}.

\subsection{Clump classification}\label{sec:classification}
The output of our CNN classifier is a value $P\!\in\![0,1]$ (produced by the sigmoid function, see \S\ref{sec:CNN}), which we interpret as the probability for a given stamp to represent a LLC, as derived by the machine. A standard way to evaluate the performance of a binary classifier is using a \textit{ROC} curve. For any chosen probability threshold $P_0\in[0,1]$, one can convert the output of the machine $P$ to a binary class $c$ by setting $c =0 $ if $P<P_0$ and $c=1$ otherwise. Then, \textit{positives} are images for which $c=1$ and \textit{negatives} are images for which $c=0$. This is compared with the true class of the image, $C$, to get the number of \textit{true positives} (TP), \textit{false positives} (FP), \textit{true negatives} (TN) and \textit{false negatives} (FN). We then define
\begin{equation}\label{eq:tpr}
    \text{recall = true positive rate (TPR)} = \frac{TP}{TP+FN},
\end{equation}
\begin{equation}\label{eq:fpr}
    \text{false positive rate (FPR)} = \frac{FP}{FP+TN},
\end{equation}
\begin{equation}\label{eq:prec}
    \text{precision} = \text{purity} = \frac{TP}{TP+FP}.
\end{equation}
By definition, $TP+FN$ is the entire positive population, and $FP+TN$ is the entire negative population. Therefore, they are independent of the threshold $P_0$ chosen. The ROC curve is a parametric curve representing the relation between TPR and FPR for varying thresholds $P_0$. One can think of TPR as being the completeness of classifying the positive class and $1\!-\!\mathrm{FPR}$ as the completeness of classifying the negative class.

The left panel of Figure~\ref{fig:roc_prec} shows an example of the resulting ROC curve for one of our classifiers. The red dots show the result for the entire validation set of this particular classifier. This model is far from being a random classifier (dashed diagonal), and it reaches a maximum completeness of $\sim 80\%$ for both classes for a probability threshold of $P_0\!\sim\!0.5$ (the knee of the curve at the top left). 

Besides completeness of the classification, we are also interested in the purity of the classification (eq. \ref{eq:prec}). This, however, is more sensitive to class imbalance. The FPs, i.e. SLCs classified as LLCs, are some fraction of the entire SLC population. Due to the large imbalance, $1\!:\!5-1\!:\!10$ between the classes, a $20\%$ misclassification of SLCs is comparable in size to the entire LLC population. Hence, $\mathrm{FP}$ can easily be comparable to $\mathrm{TP}$ in the denominator of eq~\ref{eq:prec}, making the precision very low even when only $20\%$ of the SLCs are misclassified. This is illustrated in the right panel of Figure~\ref{fig:roc_prec}, which is the \textit{precision-recall} curve, which shows the relation between $\mathrm{TPR}$ and precision. The result on the entire test set, the red curve, has precision of $\sim 50\%$ for recall of $80\%$. The thin black curves refer to precision-recall curves for randomly generated, balanced subsets from the test set. For these balanced subsets, the precision is boosted to $\sim 80\%$, as illustrated by the blue curve, showing the mean of the black curves. The same process of randomly generating balanced subsets form the test set was done for the ROC curve. The ROC curve seems to be more robust when comparing balanced and imbalanced sets. Similar results for the ROC and precision-recall curves are obtained for all the classifiers we train.

The curves in Figure~\ref{fig:roc_prec} suggest an optimal threshold $P_0$, defined as the point closest to the top-left and top-right edge of the corresponding figures, respectively. For all of our classifiers, both figures suggest an optimal threshold of $P_0\sim 0.5$, which we adopt for the rest of the analysis.

To summarize the classification performance, we are able to classify $\sim\!80\%$ of the sample, reaching purity of $\sim\!80\%$ when evaluating on balanced subsets from the test set .

\section{Clumps in the observed sample}\label{sec:observations}
\subsection{CANDELS}\label{sec:candels_intro}
Equipped with a machine trained on simulated clumpy galaxies, we now apply it to observed galaxies, aiming at detecting clumps and classifying them into LLC and SLC. For our analysis on observed galaxies, we use the CANDELS GOODS S+N fields, which are the deepest fields in the CANDELS survey (\citeb{Grogin2011}, \citeb{Koekemoer2011}).
We use a subset of galaxies used in \cite{HuertasCompany2020}. In short, the sample of galaxies consists of H-band selected objects in redshifts $1<z<3$, within the stellar mass range of $9<\log M_{\rm gal}/\Msun<12$. To avoid edge-on inclinations we select only galaxies with a 2D axial ratio $b/a>0.5$. We also mask out small galaxies with effective radii (defined as the semi-major axis of the H-band image as measured by {\tt GALFIT}) $R_e<0.2''$ ($\sim\!3$ pixels and $\sim\!1.6\ \kpc$ in the redshift range considered). Finally, we include only star-forming galaxies with specific SFR of $sSFR>10^{-10}\ {\rm yr^{-1}}$, thus eliminating galaxies that are well below the main sequence of star-forming galaxies in the relevant redshift range.
In order to have the best photometric coverage available, we only select galaxies that have images in all seven filters available. This leaves us with $\sim\!2000$ galaxies. All of the derived quantities are taken from the official CANDELS catalogs \citep{G13,Galametz2013,Santini2015,Stefano2017,Nayyeri2017,Barro2019}.

To infer physical properties of clumps, we perform SED fitting to detected clumps. To calculate the fluxes, we perform multi-channel aperture photometry with background subtraction version {\tt bgsub\_v4} as defined in \cite{G18}, and is summarized as follows. We first mask out circles of radius four pixels around each detection, to reduce the effect of blending. Then, around each detection, we calculate the median pixel value in an annulus of inner and outer radii $4<r<6$ in pixels, to calculate the background contribution to the clump. We then subtract the background contribution from the clump, to achieve the final estimate of the flux. We adopt here one successful method, namely {\tt bgsub\_v4}, and refer the reader to \cite{G18} for a comparison of different methods for background subtraction.
We then apply the Bayesian SED fitting tool {\tt bagpipes} \citep{Carnall2018} to the flux catalog. We use a $\tau$ model for the star formation history (SFH), and allow for a constant SFH as well. We use a Calzetti dust attenuation model and a Kroupa IMF \citep{Calzetti2000,Kroupa2001}. The redshift of the clumps are fixed to the redshift of the galaxy, which is either from spectroscopy or based on the best photometric redshift available. 

\smallskip The true SFH of clumps is not known. The effect of the assumed SFH on estimated parameters from the SED fitting can be significant. SFR and ages are most affected by the choice of model, and can sustain systematic offsets of $0.25- 0.3\ \mathrm{dex}$ respectively for distant galaxies, while the derived mass is usually more robust (\citeb{Lee2018}). One of the advantages of our method is that we do not assume any SFH for our classification, and depends only on the physical recipes embedded in the simulations. Since there is no reason to believe that either class suffers different systematic offsets, our method provides the opportunity to look how clump properties change between long and short lived clumps, without relying on the effects of the assumed SFH on the age.

\subsection{Results}\label{sec:results}
We apply our pipeline on all of the galaxies, to construct an observed clump catalog with positions determined by the UNet detector and longevity classification determined by our CNN classifier. We then perform the SED fitting procedure on all detected clumps to obtain the clumps' properties.
 
In order to make a fair comparison between the observed and simulated clumps, we should construct from the VELA simulations a clump catalog that undergoes the same pipeline as the observed galaxies. This is done by applying our pipeline to 17 VELA CANDELized galaxies, as described in \S\ref{sec:training_set}. This is important so both samples will be introduced to the same errors (e.g. errors introduced by the machine, SED fitting errors etc.). A comparison of the 3D clump catalog in VELA with the VELA CANDELized clumps and with the CANDELS clump properties has been carried out in \cite{HuertasCompany2020}. Here, we refer to VELA's clumps only via the CANDELized dataset. Furthermore, given the different distribution of galaxy masses, redshifts and galaxy sizes between VELA and CANDELS, we randomly draw CANDELS galaxies to roughly match the VELA distribution.

In order to reduce contamination by the bulge, we only analyze clumps detected outside $0.5R_{\rm e}$. Also, since in this study we choose to focus on in-situ clumps, we only analyze clumps within $3R_{\rm e}$ in an attempt to reduce contamination by merging galaxies. Similar constraints were applied in other analyses \citep{Shibuya2016,G18,HuertasCompany2020}.
\begin{figure}
    \centering
    \includegraphics[width=\columnwidth]{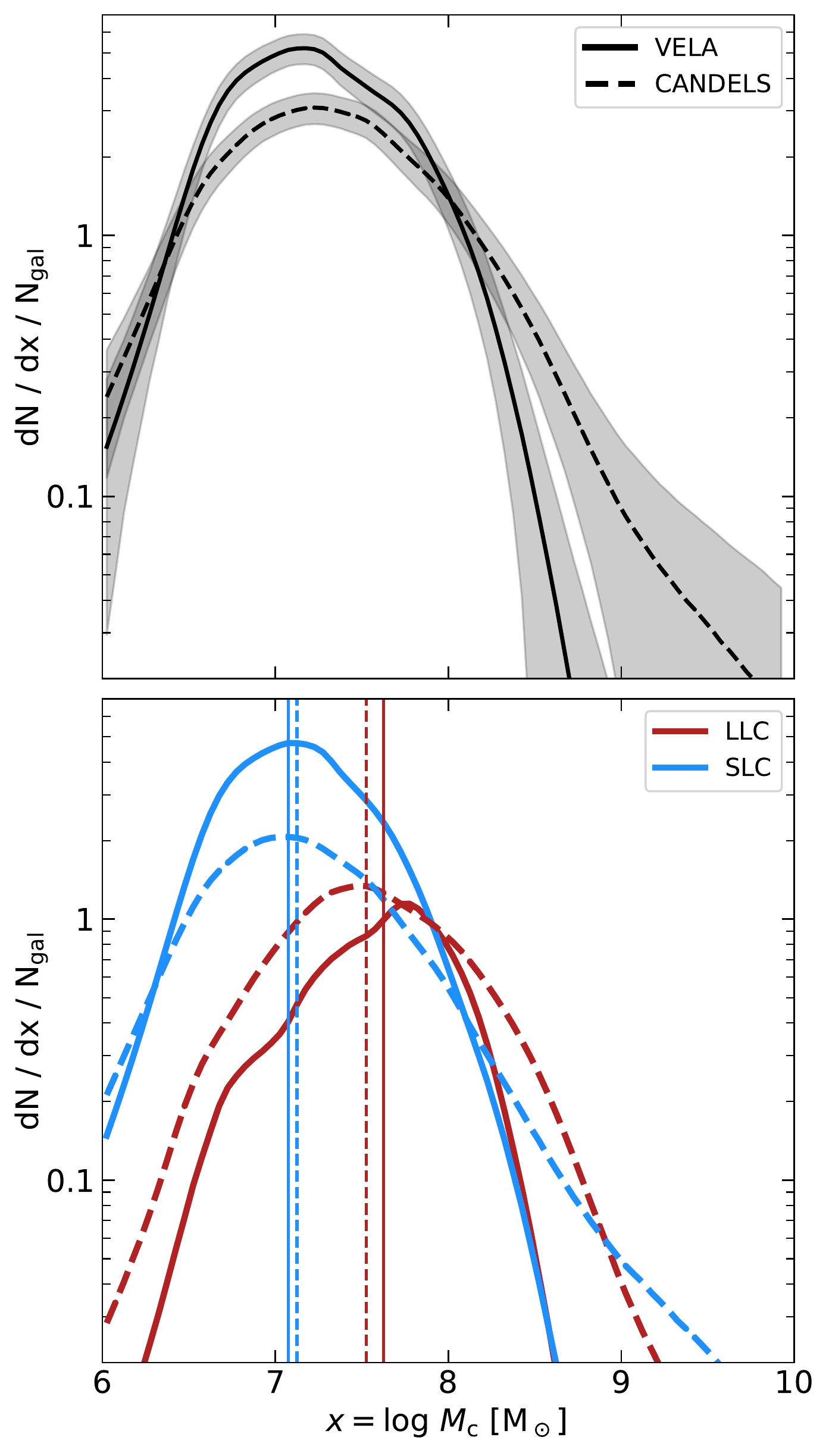}
    \caption{The clump mass function - number of clumps per logarithmic clump-mass bin per galaxy. Masses in both panels are corrected using the~\protect\cite{HuertasCompany2020}~scheme with the VELA prior (see text). Only clumps that satisfy $M_{\rm c}/M_{\rm gal} < 0.2$  are included in the analysis, in order to minimize the contamination by ex-situ merged clumps. \textbf{\textit{Top}}: The total clump mass functions of VELA and CANDELS. The solid curve refers to the mass function of VELA clumps, while dashed curve refers to the mass function of CANDELS clumps.  \textbf{\textit{Bottom}}: Clump mass function of each population (red for LLCs, blue for SLCs). The vertical lines refer to the median of each population. We can see that LLCs tend to be more massive than SLCs, both in VELA and in CANDELS, and that there is a general agreement between the number of clumps per galaxy for each type in each sample.}
    \label{fig:cmf}
\end{figure}

\subsubsection{Mass functions for LLC and SLC in CANDELS vs VELA}
Figure~\ref{fig:cmf} shows the clump mass function of the clumps detected by our deep learning pipeline, using the masses derived from the SED fitting scheme. \cite{HuertasCompany2020} studied the observational effects on clump masses estimated using SED fitting, and found a systematic offset (with respect to the 3D clumps in VELA) of up to $\sim\!\!1\ \dex$, with a secondary dependence on the distance from the centre. Here, we apply to the SED-fitted masses the Bayesian correction scheme proposed by \cite{HuertasCompany2020} for a better match to the masses in the simulations. \cite{HuertasCompany2020} suggested two different priors for correcting the masses. Here, we use the `VELA' prior, which uses the 3D VELA clump mass function as the prior. This prior is biased at the poorly populated tails of the mass function, at $M_{\rm c}\lesssim10^6\ \Msun$ and at $M_{\rm c}\gtrsim10^9\ \Msun$. It should be noted that the other prior proposed by \cite{HuertasCompany2020} has a flat massive-end tail, emphasizing the uncertainty in the masses of massive clumps. The reader is referred to that paper for more details and comparison between the methods. 

From the top panel of Figure~\ref{fig:cmf}, we read that below $10^8\ \Msun$, the clump mass function of VELA is higher than that of CANDELS, by up to $0.3\ \dex$, while at higher masses the mass function of CANDELS is higher. The log slope of the mass function at the high mass range ($M_{\rm c}>10^{7.5}$) is $-1.87\pm 0.24$ for VELA and $-0.86 \pm 0.21$ for CANDELS.
The shallower slope in CANDELS may be due to ex-situ clumps. We tried to reduce the effect of ex-situ clumps by excluding very massive clumps with $M_{\rm c}> 0.2 M_{\rm gal}$, but this does not eliminate the whole effect. VELA galaxies are less susceptible to being contaminated by ex-situ clumps, as they are chosen to be rather isolated (see \S\ref{sec:vela}).

The results in the top panel of Figure~\ref{fig:cmf} are in overall agreement with \cite{HuertasCompany2020}, which found that CANDELS clump mass function is higher than VELA in the entire mass range. In our case, the CANDELS mass function is higher than the VELA mass function only above $10^8\ \Msun$. This is because the similar methods used in the two papers have two important differences. First, in \cite{HuertasCompany2020} we detected clumps in only one band, either rest-frame optical or rest-frame UV, while here we use seven bands. Second, in \cite{HuertasCompany2020} we trained the UNet with mock galaxies images. These are 2D S\'{e}rsic light distributions with manually put clumps, while here we train the network with mock images of simulated galaxies, which are more complicated structures.

The bottom panel of Figure~\ref{fig:cmf} shows the mass functions of LLCs and SLCs in the two samples. First, we see in both samples that the LLCs tend to be more massive than the SLCs, with a $\sim 0.5\ \dex$ difference between the medians of the populations. We can see that VELA contains more SLCs than CANDELS in the clump mass regime of $M_c\lesssim 10^8\ \Msun$. On the other hand, CANDELS contains more LLCs than VELA by a factor of $\sim2$ in the low mass regime (i.e. $M_{\rm c}\lesssim10^{8}\ \Msun$), and the difference at the high mass end is even larger. However, as mentioned above, the inability to distinguish ex-situ and in-situ clumps may be the cause for the shallower massive-end slope in CANDELS.

Another feature visible in the bottom panel of Figure~\ref{fig:cmf} is that CANDELS show a transition mass, $M_{\rm c}\sim 10^{7.6}\ \Msun$, above which most clumps are LLCs. VELA shows a similar transition at, around $M_{\rm c}\sim 10^{7.9}\ \Msun$. This difference is of marginal significance.

Recall that during training, we remove the mass information by randomizing the luminosities of the trained galaxies and clumps (see \S\ref{sec:cnn_training} above). The fact that, in spite of this procedure, our pipeline finds that long lived clumps are more massive, indicates that our machine indeed learns other properties of the clumps, beyond the mass.
Overall, we can see that the typical masses of each clump population in each sample agree to within $10-20\%$ in medians, and the mass functions agree within a factor of two in the mass regime of $10^{7-8}\ \Msun$. Given the elimination of the explicit information concerning the clump masses from the training set, this agreement is a meaningful, encouraging non-trivial result.

\begin{figure*}
    \centering
    \includegraphics[width=\columnwidth]{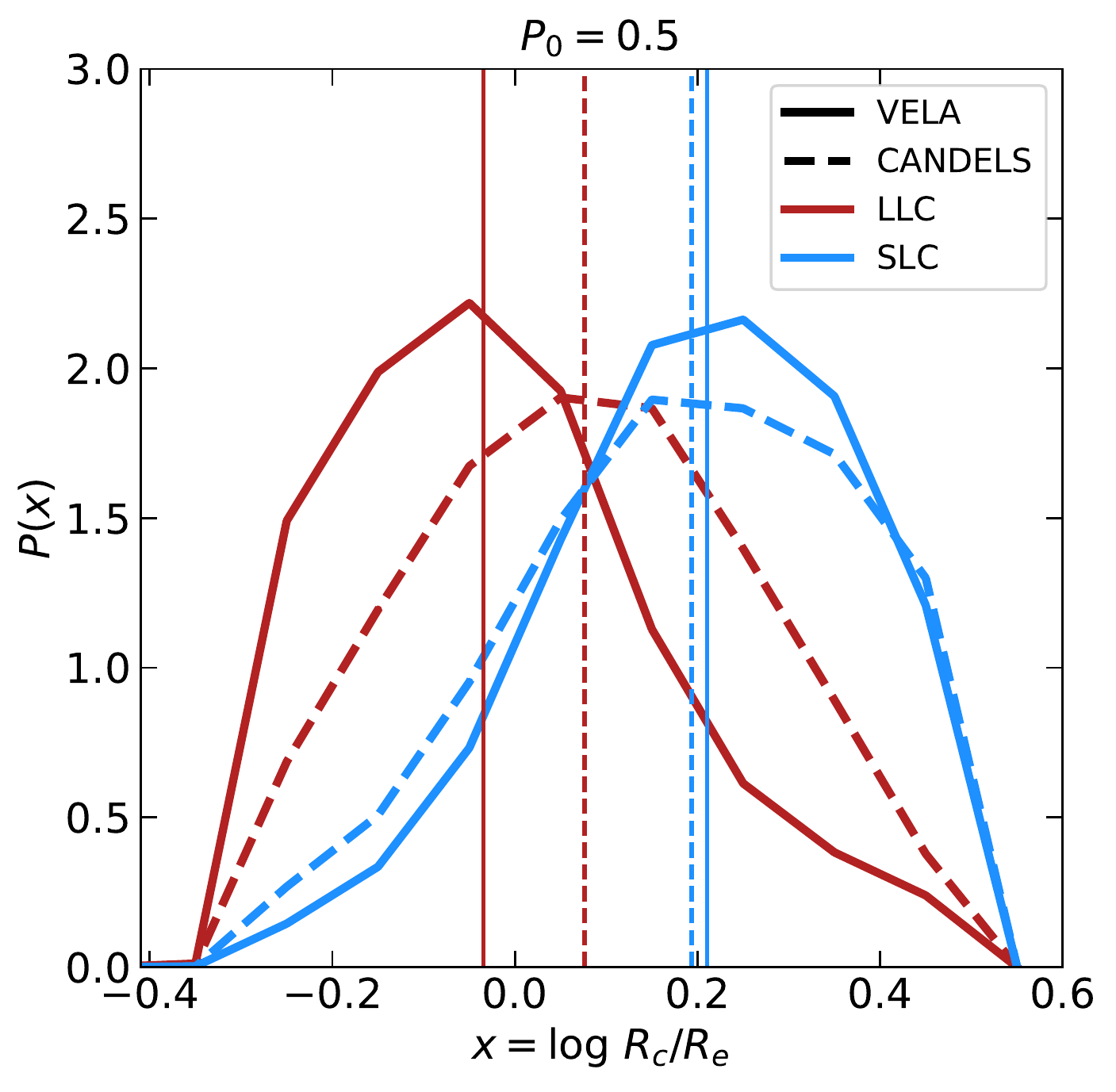}
    \includegraphics[width=\columnwidth]{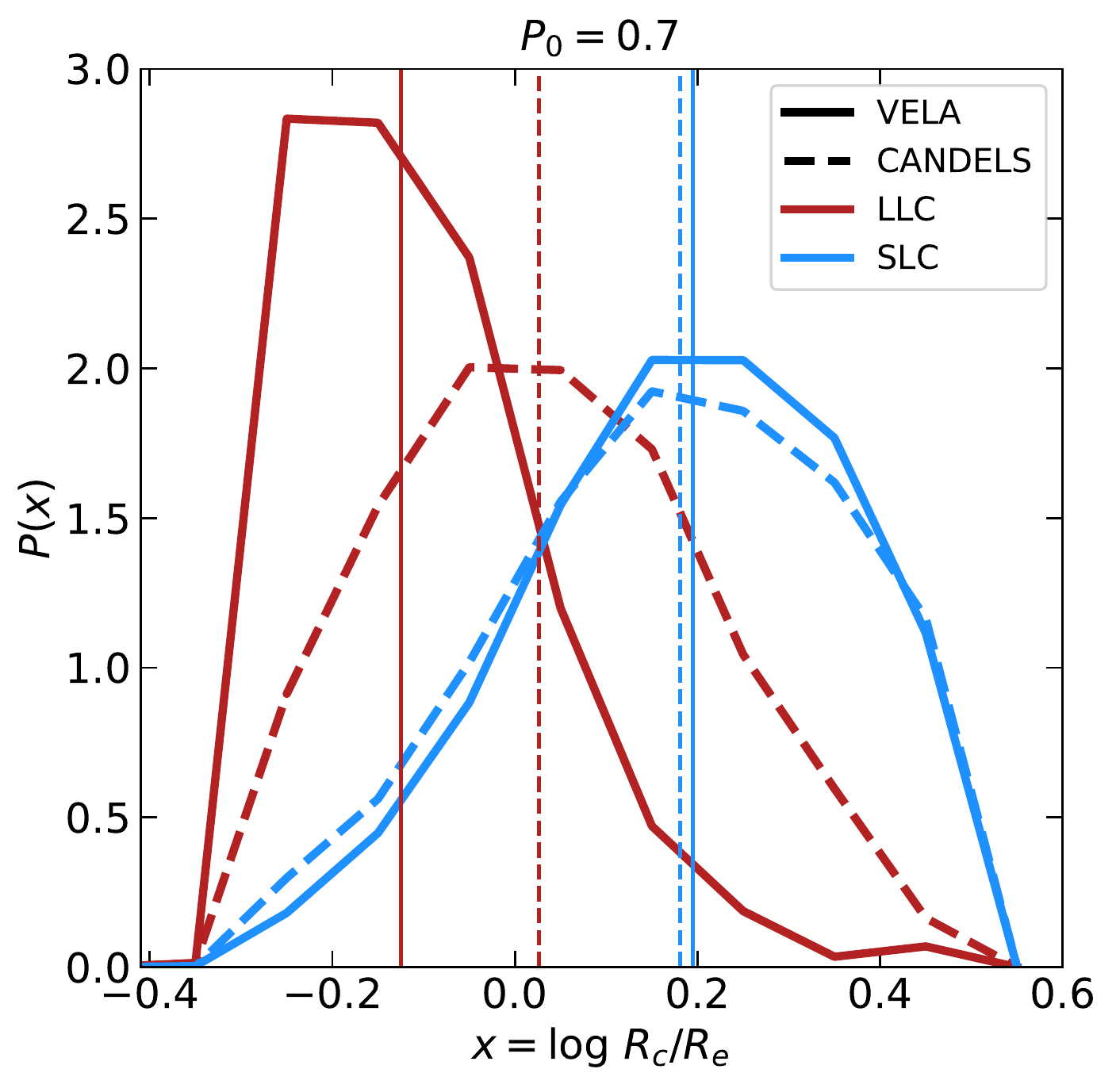}
    \caption{Probability distributions of clump radial position (in terms of $R_{\rm e}$), separated to SLC and LLC in both VELA and CANDELS (see labels). Vertical lines indicate the medians of each distribution. The clump classes are determined by a probability threshold on the machine's output of $P_0=0.5$ and $0.7$ in the left and right panels, respectively. We can see that regardless of sample and threshold, the LLCs are generally found to be closer than SLCs to the centre. The LLCs are closer to the centre in VELA than in CANDELS}
    \label{fig:migration}
\end{figure*}
\subsubsection{Clump migration}
In order to address the possibility of LLC migration, we explore the radial positions of the clumps with respect to the galactic centre. Figure~\ref{fig:migration} shows the probability distribution of the clumps' radial position, normalized by the (two dimensional) effective radius of the galaxy, for all clumps, separated into SLCs and LLCs, in VELA and in CANDELS.

The left-hand panel shows the distributions for the fiducial probability threshold (on the machine's output) of $P_0=0.5$. For this threshold, $\sim\!40\%$ of LLCs in CANDELS and $\sim\!50\%$ of the LLCs in VELA reside within $R_{\rm e}$.
The difference between the median positions of SLCs and LLCs in VELA is $\sim 0.2-0.3\ \dex$, roughly consistent with \cite{M17} (e.g. Figure 15 therein). The difference between the corresponding medians in CANDELS is somewhat smaller, $0.12-0.15\ \dex$. This is consistent with the age gradients found in \cite{G18} (e.g. Figure 9 therein). If we roughly set there an age threshold of $300\ {\rm Myr }$ to determine whether a clump is long or short lived (see Figure~\ref{fig:age_color} below), we read from Figure 9 in \cite{G18} a $\sim 0.1\ {\rm dex}$ difference between the median distance of the younger ($<300\ \Myr$) compared to the older ($>300\ \Myr$) clumps. 

We note that, while the distribution of SLC positions in VELA and CANDELS are in good agreement, VELA LLCs are usually found closer to the centre than CANDELS LLCs. The right panel in Figure~\ref{fig:migration} shows the distribution for a probability threshold of ${P_0 = 0.7}$, which corresponds to a cleaner, less complete sample. While the distribution of SLC positions remains unchanged, LLCs detected with this higher threshold reside closer to the centre.

Interpreting the difference in positions of SLCs and LLCs as being due to clump migration, the results of Figure~\ref{fig:migration} suggest that clumps in VELA tend to migrate more efficiently than in CANDELS galaxies. This could be due to the relatively weak feedback implemented in VELA, which allows the LLCs to be more compact and hence more resistant to tides and disruption during the migration.
\begin{figure}
    \centering
    \includegraphics[width=\columnwidth]{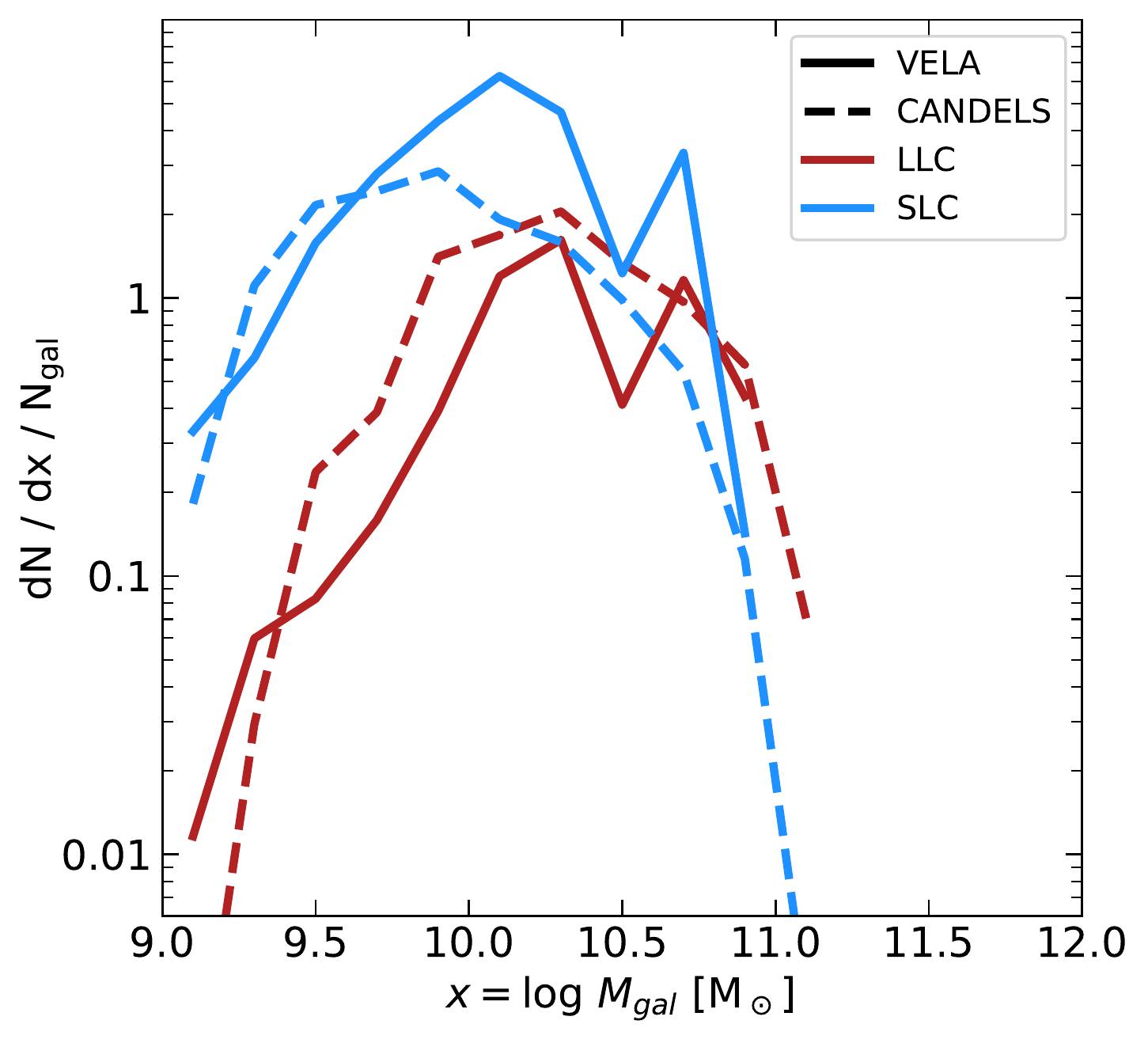}
    \caption{Distribution of clumps' host galaxy stellar mass (see labels). Each galaxy is counted as many times as there are clumps of the given type detected in that galaxy. LLCs tend to emerge in high mass galaxies, of $10^{10}\ \Msun$ and above. CANDELS shows a transition between SLC dominance and LLC dominance at $M_{\rm gal}\!\sim\!10^{10.2}$, while in VELA the transition is at higher masses.}
    \label{fig:clump_mgal}
\end{figure}

The difference in positions of SLCs and LLCs in Figure~\ref{fig:migration} may, alternatively, imply that LLCs preferentially form closer to the centre. \cite{M17} examined the gradient of the clump time, $t_{\rm c}$ (see \S\ref{sec:in_situ} above), of SLCs and LLCs in VELA. They found that newly formed clumps, with small $t_{\rm c}$, are found at the outskirts of the disc, regardless of their longevity class (see Figure 14 therein). This hints that LLCs are formed, together with SLCs, at the outskirts of the galaxies and are not preferentially formed in the innermost parts. However, a more detailed study, both analytical and using simulations, is needed, which we leave for future work.

\subsubsection{Host galaxy connection}
Figure~\ref{fig:clump_mgal} shows the distribution of clumps' host galaxy mass, for the two clump populations, in VELA and in CANDELS. It is evident that LLCs mainly appear in galaxies with mass $\gtrsim\!10^{10}\ \Msun$, and that the average number of SLCs start to decrease above this mass, in both VELA and CANDELS. CANDELS exhibits a transition between galaxies hosting mostly SLCs to LLCs at a galaxy mass of $M_{\rm gal}\!\sim\!10^{10.2} \Msun$.
VELA shows a hint for such transition at a higher mass of $M_{\rm gal}\!\sim\!10^{10.8}\ \Msun$. 
 
According to the theory of clump formation due to Toomre instability during VDI, the clump mass is proportional to the disc mass \citep*{DekelSariCeverino09}. Thus, more massive galaxies are expected to host more massive clumps, which are more likely to survive the effects of stellar and supernova feedback. This would result in more massive galaxies being more likely to host LLCs. More on the effects of supernova feedback on clump survival in \S\ref{sec:spnv}.
\section{Discussion}\label{sec:discussion}
\subsection{What does the machine use for clump classification?}
Even though deep NNs are considered black boxes, it is desirable to gain some understanding of the key drivers of the machine decisions. While a systematic analysis of this sort is beyond the scope of the current paper, a conceptual understanding is not out of reach. In the VELA simulations, we have access to many resolved quantities that are not available in observations. In particular, there are several physical parameters that correlate with the definition of long-lived clumps, which may dominate the machine's decision. The most straightforward parameters are clump mass and age. Long lived clumps form stars for a longer period of time, and, for a fixed clump radius, more massive clumps are expected to be more resistant to disruption. While completely or partially unresolved in observations, shape and density contrast are strong indicators in VELA for whether a clump is short lived or long lived. Long lived clumps are rounder (with shape parameter $S_{\rm c}\gtrsim 0.8$, see \S\ref{sec:clump_finder} for a definition) than short-lived clumps ($S_{\rm c}\lesssim0.4$, see Figure~\ref{fig:zlc_dist}), and they have a higher density contrast, with respect to their surroundings. This is one of the properties that make them long lived. In fact, we find in VELA that in the first snapshot in which they are identified, long lived clumps already have a high density contrast.

What does our machine use for classification? Since the mass information, which corresponds to the absolute value of the pixels, is hidden from the machine during training, by randomizing the magnitudes (see \S\ref{sec:training_set}), the only other information available to the machine is color information (the relative fluxes between the different filters) and structural information. To test the effect of the color of the clumps, we completely randomized the magnitudes in each filter independently, without changing the density distribution that tells the structure within each image. For the same network architectures used in the main study, the machine performance became much inferior, reaching a completeness of $\sim\!50\%$ and a purity of $\sim\!40\%$, as compared to the $\sim 80\%$ achieved in both scores by the fiducial models. The structural information of the clump is probably less accessible to the machine, due to the small square size we are using and the limited resolution.

\begin{figure}
    \centering
    \includegraphics[width=\columnwidth]{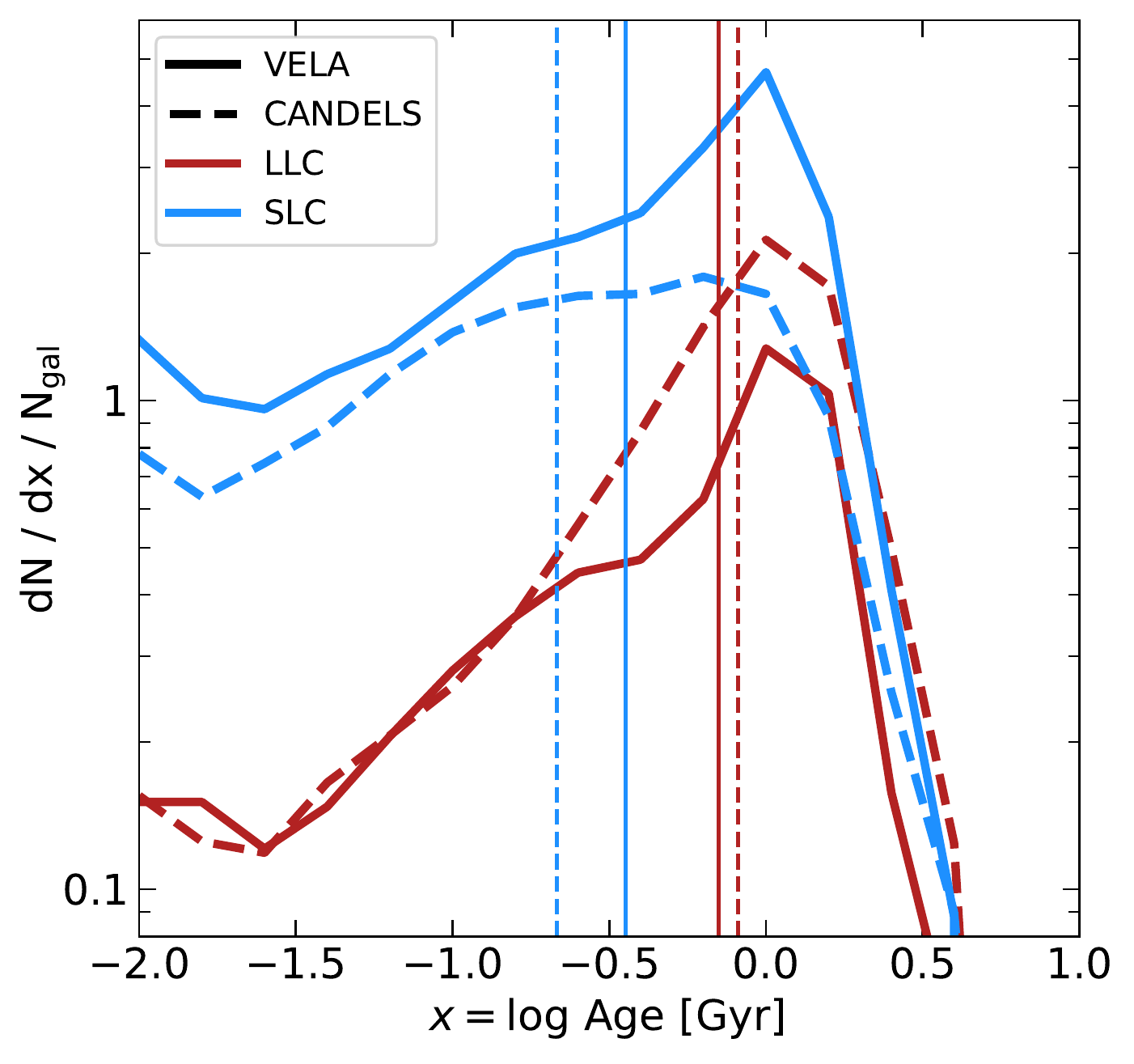}
    \caption{Distribution of clump ages in both CANDELS (dashed) and VELA CANDELized (solid), separated to different longevity classes. The vertical lines are the medians of each distribution. As expected in this sanity-check, clumps classified as LLCs are mostly older, while the SLCs tend to be younger. The high mass tail of the SLC distribution is largely due to contamination by background stars (see text).}
    \label{fig:age_color}
\end{figure}

To even further investigate this, we examine whether or not there is a correlation between our classification to SLCs and LLCs and age, estimated from the SED fitting. Figure~\ref{fig:age_color} shows the distribution of ages of both VELA and CANDELS clumps of the two types, as estimated from the SED fitting. We can see that LLCs tend to be older, with a median age of $\sim 900\ \rm{Myr}$, compared to only $300\ \Myr$ for SLCs. We do note a high mass tail in the SLC distribution. This is largely due to the contamination by background stars. Indeed, as discussed in \cite{G18}, the ages and colors of clumps at the outskirts of the galaxy, which tend to be short lived (see Fig. \ref{fig:migration}), are more affected by background stars.

It is important to note that the method we have developed in this study does not assume any underlying model for deducing clump properties from the observed fluxes, as opposed to the different models used in the SED fitting. Rather, they only rely on the subgrid models implemented in our simulations, and the models used to generate fluxes for the mock images. However, the SED fitting procedure is performed independently, and the fact that our machine's predictions and the SED-derived properties correlate, suggests that our machine is able to learn the underlying physical quantities that are associated with the clump's longevity, beyond just the clump mass.

\subsection{The effect of supernova feedback on clump longevity}\label{sec:spnv}
Why do higher mass galaxies' clumps live longer? \cite*{DekelSariCeverino09} derived an analytical condition for the disruption of clumps due to supernova feedback, based on \cite{DekelSilk86}. The key parameter is the clump's potential well depth, characterized by the circular velocity $V_{\rm c,circ}$. The threshold for clump survival was found to be
$$V_{\rm SN,c}\approx 30\ {\rm km\cdot s^{-1}}\ \epsilon_{0.1}^{1/2}$$ 
where $\epsilon_{0.1} = \epsilon/0.1$, $\epsilon$ is the star formation efficiency, $\epsilon = t_{\rm ff}/t_{\rm dep}$, with $t_{\rm ff} = \sqrt{3\pi/32G\rho}, t_{\rm dep}=M_{\rm gas}/{\rm SFR}$ as the free fall and depletion times, respectively. \cite*{DekelSariCeverino09} derived an effective value of $\epsilon\sim0.06-0.1$. \cite{M17} found that when averaged over clumps in VELA, the SFR efficiency is typically $\epsilon\sim 0.03-0.04$. These values are larger than the observed star formation efficiencies in star-forming clouds \citep{Utomo2018} and analytic estimates \citep{Krumholz2005}, which find $\epsilon\sim0.01$. \cite*{DekelSariCeverino09} suggested that this discrepancy is due to star formation in clumps that actually occurs on smaller, unresolved scales within the clumps, each with efficiency of $0.01$.

The circular velocity of giant clumps can be estimated using eqs. 8,9 in \cite*{DekelSariCeverino09}

\begin{equation}
\centering
    V_{\rm circ,c} \approx \sqrt{\frac{\pi}{6}}\delta_{\rm d}V_{\rm circ,gal} = \sqrt{\frac{\pi}{6}}\delta_{\rm d}f_{\rm vir}V_{\rm vir}.
\end{equation}
where $\delta_{\rm d} = M_{\rm cold}/M_{\rm tot}$ is the ratio between the cold mass in the disc and the total mass within the disc radius, $V_{\rm circ,gal}$ is the galaxy's circular velocity, $V_{\rm vir}$ is the virial velocity of the galaxy's host halo and $f_{\rm vir}=V_{\rm circ,gal}/V_{\rm vir}\sim 1.4$ (e.g. Figure 7 in \citeb{DekelDisk}). Assuming fiducial values of $\delta_{\rm d} = 0.2, f_{\rm vir}=1.4$ \citep*[e.g.][]{DekelSariCeverino09, DekelRings}, we get
\begin{equation}
    V_{\rm circ,c} \approx 0.2\cdot \delta_{\rm d,0.2}f_{\rm vir,1.4}V_{\rm vir}.
\end{equation}
where $\delta_{\rm d,0.2}=\delta_{\rm d}/0.2,\ f_{\rm vir,1.4} = f_{\rm vir}/1.4$. The threshold for disruption of clumps by supernova feedback is therefore translated to a threshold on $V_{\rm vir}$ of the galaxy's host halo
\begin{equation}\label{eq:sn_thresh}
    V_{\rm vir,SN,c} \approx 150\ {\rm km\cdot s^{-1}}\epsilon_{0.1}^{1/2}\delta_{\rm d,0.2}^{-1}f_{\rm vir,1.4}^{-1}.
\end{equation}
Interestingly, this threshold is in the ballpark of the \cite{DekelSilk86} threshold for supernova efficiency on a global scale. While $f_{\rm vir}$ is usually of order unity, in the range of $1-1.5$, $\delta_{\rm d}$ can easily vary by a factor of $\sim 2$ from our fiducial value (e.g. \citeb{DekelRings}). 

From standard cosmology, the threshold in equation \ref{eq:sn_thresh} can be translated to a threshold on halo mass. In the Einstein-deSitter cosmology regime, appropriate for $z\gtrsim 1$, we can approximate (e.g. \citeb{Dekel2013})
\begin{equation}
    M_{\rm v} \approx 10^{12}\ \Msun\cdot \lbrac{\frac{V_{\rm vir}}{200\ {\rm km\cdot s^{-1}}}}^3\lbrac{\frac{1+z}{3}}^{-3/2}.
\end{equation}
For $V_{\rm vir} = V_{\rm vir,SN,c}$, this yields
\begin{equation}\label{eq:m_sn_thresh}
    M_{\rm v,SN,c} \approx 4.2\cdot 10^{11}\ \Msun\ \epsilon_{0.1}^{3/2}\delta_{\rm d,0.2}^{-3}f_{\rm vir,1.4}^{-3}\lbrac{\frac{1+z}{3}}^{-3/2}.
\end{equation}
This ranges from $2.7\times 10^{11}\ \Msun$ to $7.7\times 10^{11}\ \Msun$ in the redshift range of $1<\!z\!<3$. We notice the very strong dependence on $\delta_{\rm d}$, which has the potential to increase this threshold mass by an order of magnitude.

Using the empirical relation between $M_{\rm v}$ and $M_{\rm gal}$ proposed by \cite{Behroozi19}, the threshold on the halo mass can be translated to a threshold on the galaxy mass, and for our fiducial values, the threshold stellar mass is in the range ${M_{\rm gal}\sim2\times10^{9}\ \Msun-1.6\times10^{10}\ \Msun}$, roughly consistent with the results in Figure~\ref{fig:clump_mgal}.

\section{Conclusions}\label{sec:conc}
We used deep learning techniques to study the nature of clumps in massive, star-forming disc galaxies at $z\sim1-3$. Using the 3D clump catalog from the VELA high resolution zoom-in simulations (\citeb{M17}), we trained a deep learning pipeline to detect clumps in HST-like mock images of the simulations and then classified them according to their appropriate longevity class, either short lived or long lived clumps (SLCs or LLCs). Our pipeline's detector is able to detect $\sim80\%$ of the simulated clumps in terms of mass (above $\sim\!10^{7.5-8}$ for baryonic and stellar masses, respectively) over the whole studied redshift range, with $\sim\!80\%$ purity rates. Our pipeline's classifier is able to classify correctly $\sim\!80\%$ of our simulated clumps, with $\sim\!80\%$ purity rate when estimating on balanced sets.

We then ran observed disc galaxies from CANDELS GOODS S+N through our pipeline to detect and classify the clumps, and compare the properties with the simulated clumps. We summarize our findings as follows.
\begin{itemize}
    \item The clump mass functions of VELA and CANDELS agree to within a factor of $\sim 2$ in the clump mass regime of $M_{\rm c}\lesssim10^{8}\ \Msun$. Above this mass, the mass function of CANDELS clumps is shallower and higher than VELA, possibly due to ex-situ clumps.
    \smallskip\item VELA galaxies contain a factor of $\sim 2$ more SLCs than CANDELS in the mass regime $\lesssim10^{8}\ \Msun$, and CANDELS galaxies contain a factor of $\sim2$ more LLCs in this mass regime.
    \smallskip\item The LLCs tend to be more massive than SLCs by $0.5\ \dex$. The medians of the mass functions of the SLCs and LLCs are $10^{7}\ \Msun$ and $10^{7.5}\ \Msun$, respectively, with good agreement between CANDELS and VELA.
    \smallskip\item When inspecting the mass functions of SLCs and LLCs, we find a transition mass of $\sim10^{7.6-7.9}\ \Msun$ in both CANDELS and VELA, respectively. Above this mass, most clumps are LLCs.
    \smallskip\item LLCs tend to reside closer to the galactic centre, within or near the effective radius in both VELA and CANDELS. This result can be interpreted as migration of LLCs to the centre of the galaxy. In this scenario, it appears that the migration is stronger in VELA than in CANDELS, possibly indicating an underestimate of the feedback strength. The proximity to the centre may also be interpreted as a preferred formation location of LLCs,but this is not supported by our finding in the simulations \citep[][Figure 14]{M17}.
    \smallskip\item LLCs tend to reside in high mass galaxies, and they become dominant in galaxies with masses above $10^{10.2}\ \Msun$ and $10^{10.8}\ \Msun$ in CANDELS and VELA, respectively. This can be explained using the survival of clumps under supernova feedback in high mass galaxies. In such galaxies, clumps are more massive and hence with a deeper potential well.
    \smallskip\item Our machine classification correlates with the stellar age as estimated from SED fitting, which is performed independently, with median SLC age $\sim 300\ \Myr$ and median LLC age $\sim 900\ \Myr$.
\end{itemize}

In this study, even though we pushed the limits of conventional deep learning practices to small samples, we were able to construct an efficient tool to compare theoretical models of clump formation in a cosmological environment to real observed galaxies. 
To improve the quality of the deep learning pipeline for a cleaner comparison between simulations and observations, one needs a larger dataset. This is rather demanding because of the high resolution of tens of parsecs that is required for a proper study of clumps. 

The results of this paper are for the VELA3 simulation suite, whose incorporated feedback recipes may be on the weak side globally, based on the stellar-to-halo mass ratios. Nevertheless, the agreements found here with CANDELS galaxies is an indication that the feedback, as implemented, is in the appropriate ballpark on the clump scale. Preliminary inspection of VELA6, a new re-simulation of VELA3 galaxies, which incorporate a globally stronger feedback with the rest of the sub-grid physics kept intact (Ceverino et al., in prep.), indicates that the number of simulated long lived clumps is smaller. Future studies should construct a similar pipeline based on simulations with different feedback recipes, in which  either LLCs are more abundant \citep[e.g.][which also studied clumps in VELA without radiation pressure]{M17}, or in which most clumps are short lived \citep[e.g.][]{Genel12,Oklopcic2017}, in order to see if our results change based on the underlying dataset.

\section*{Acknowledgements}
We acknowledge helpful discussions with Ryan Hausen regarding the UNet. We also acknowledge Christoph Lee for the initial work on clump detection using the UNet. This work was partly supported by the grants DIP STE1869/2-1 GE625/17-1 and ISF 861/20. NM acknowledges support from the Gordon and Betty Moore Foundation through Grant GBMF7392, and from the Klauss Tschira Foundation through the HITS Yale Program in Astrophysics (HYPA). DC is a Ramon-Cajal Researcher and is supported by the Ministerio de Ciencia, Innovaci\'{o}n y Universidades (MICIU/FEDER) under research grant PGC2018-094975-C21. JP acknowledges support from Google Faculty Research Grant and from grant HST-AR-14578.001-A from the STScI under NASA contract NAS5-26555. The VELA simulations were performed at the National Energy Research Scientific Computing Center (NERSC) at Lawrence Berkeley National Laboratory, and at NASA Advanced Supercomputing (NAS) at NASA Ames Research Center. Development and analysis of the simulations have been performed in the Astro cluster at HU. The VELA CANDELized images were generated at NASA Ames Research Center.

\section*{Data Availability}
The VELA CANDELized images are publicly available at \href{https://archive.stsci.edu/prepds/vela/}{https://archive.stsci.edu/prepds/vela/}.
The data underlying this article, namely the simulation data, clump catalogs and machine-produced catalogs, will be shared on reasonable request to the corresponding author, requiring the consent of all of the co-authors.




\bibliographystyle{mnras}
\bibliography{example} 





\bsp	
\label{lastpage}
\end{document}